\documentclass[journal=jpcbfk,manuscript=article,layout=traditional]{achemso}

\usepackage{hyperref}
\usepackage{verbatim}
\usepackage{amsmath}
\usepackage{amssymb}
\usepackage{graphicx}   % Include figure files
\usepackage{caption}
\usepackage{subcaption}
\usepackage{dcolumn}    % Align table columns on decimal point
\usepackage{bm}         % bold math
\usepackage{xr}
\usepackage[version=3]{mhchem} % Formula subscripts
\usepackage{xcolor}

\makeatletter
\def\acs@author@fnsymbol#1{}
\makeatother

\author{Thomas Petersen$^\mathrm{\dagger}$}
\affiliation[University of Southern California]{$^\mathrm{\dagger}$Sonny Astani Department of Civil and Environmental Engineering\\ University of Southern California, Los Angeles, California 90089 USA}
\email{thomasp3@usc.edu}
\phone{+1 (928) 210-2088}

\title{Toward Modeling the Structure of Electrolytes at Charged Mineral Interfaces using Classical Density Functional Theory}

\begin{document}

\begin{tocentry}
\includegraphics{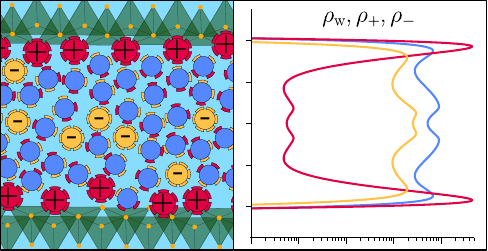}
\end{tocentry}

\begin{abstract}
The organization of water molecules and ions between charged mineral surfaces determines the stability of colloidal suspensions and the strength of phase-separated particulate gels. In this article we assemble a density functional that measures the free energy due to the interaction of water molecules and ions in electric double layers. The model accounts for the finite size of the particles using fundamental measure theory, hydrogen-bonding between water molecules using Wertheim's statistical association theory, long-range dispersion interactions using Barker and Henderson's high temperature expansion, electrostatic correlations using a functionalized mean-spherical approximation, and Coulomb forces through the Poisson equation. These contributions are shown to produce highly correlated structures, aptly rendering the layering of counter-ions and co-ions at highly charged surfaces, and permitting the solvation of ions and surfaces to be measured by a combination of short-ranged associations and long-ranged attractions. The model is tested in a planar geometry near soft, charged surfaces to reproduce the structure of water near graphene and mica. For mica surfaces, explicitly representing the density of the outer oxygen layer of the exposed silica tetrahedra allows water molecules to hydrogen-bond to the solid. When electrostatic interactions are included, water molecules assume a hybrid character, being accounted for implicitly in the dielectric constant but explicitly otherwise. The disjoining pressure between approaching like-charged surfaces is calculated, demonstrating the model's ability to probe pressure oscillations that arise during the expulsion of ions and water layers from the interfacial gap, and predict strong inter-attractive stresses that form at narrow interfacial spacing when the surface charge is overscreened. This inter-attractive stress arises not due to in-plane correlations under strong electrostatic coupling, but due to out-of-plane structuring of associating ions and water molecules.
\end{abstract}

\section{Introduction}

Despite over a century of theoretical developments since the introduction of the Poisson-Boltzmann equation~\cite{gouy1910constitution,chapman1913li}, modeling ion concentrations within electric double layers (EDLs) remains a significant challenge, particularly where the bulk salt concentration is large and multivalent ions are present. The measurement of ion distributions near interfaces is difficult as the screening length typically only extends a few nanometers -- the Stern layer in aqueous electrolytes, for instance, extends only 1-2 water monolayers ($\sim6$\r{A}) -- and many materials of industrial interest display a surface texture that may be more or less ordered. Calcium-silicate-hydrate (C-S-H), the principal binding phase in cement, for instance, possesses a semi-crystalline, layered molecular structure that can vary as a function of the ratio of calcium to silica or the substitution of calcium for other cations~\cite{pellenq2009realistic,skibsted1994direct}. Nonetheless, the structure of ions and water molecules near C-S-H's charged silica layers is known to bestow it its binding properties~\cite{pellenq1997electrostatic,goyal2021physics}. Electrochemical double layer capacitors store charge by adsorbing ions in high-surface-area porous electrodes~\cite{chmiola2006anomalous}; for carbide-derived carbon supercapacitors it is now known that the ion exchange rate at the electrode depends significantly on the pore size and microstructure~\cite{merlet2012molecular}. Even muscovite mica, which is often chosen as a model material for surface force studies due to its near-atomically flat surface, contains interstices between silica tetrahedra that permit adsorption of small ions and water molecules, which affects the local liquid structure~\cite{park2002structure,park2006hydration}. Thus, scientists often rely on measurement of macroscopic properties, such as surface disjoining pressures~\cite{pashley1984molecular,plassard2005nanoscale} and surface or zeta potentials~\cite{lutzenkirchen2012potentiometric}, to make statements about the construction of the liquid-solid interface. Though, we note, advances in electromagnetic and optical analytical techniques are making headway in direct measurement~\cite{zaera2012probing}.

An additional challenge is the measurement of electrostatic correlations, which are not captured by the mean-field Poisson-Boltzmann theory. They lend important effects to the layering and solvation of ions and interfaces~\cite{goyal2021physics,lee2021ion} and can lead to attraction of like-charge surfaces~\cite{pellenq1997electrostatic,plassard2005nanoscale,roth2016shells}, charge-reversal of colloidal particles~\cite{torrie1980electrical} and nanochannels~\cite{qiao2004charge}, or assist with conduction of ions through biological membranes~\cite{liu2013correlated}. Many field theoretic models exist to capture electrostatic correlations~\cite{bazant2011double,kierlik1991density,gillespie2002coupling,gupta2020ionic}, though they are often not coupled to other important forces that need to be included when modeling highly concentrated electrolytes or electrolytes at highly charged interfaces. For instance, excluded volume interactions are required to properly model layering of ions in room-temperature ionic liquids~\cite{de2020interfacial} or the oscillatory disjoining pressures that correspond to the displacement of layered water molecules in approaching mica surfaces~\cite{pashley1984molecular}. Furthermore, accurate prediction of thermodynamic parameters requires the energetics of the ion-water interactions -- from London dispersion interactions to hydrogen bonding -- to be properly measured~\cite{chapman1989saft}.

In this article, we sample several theories to model the structure of water and ions near textured mineral interfaces using classical density functional theory (cDFT). Equilibrium cDFT is utilized to model thermodynamically averaged density distributions of water molecules and ions, providing an efficient approach to modeling the inhomogeneous structure of electrochemical fluids near interfaces that are otherwise only accessible through computationally intensive Monte Carlo (MC) or molecular dynamics (MD) simulations~\cite{hansen2013theory}. The model is built step-by-step to match the density distribution of water molecules near graphene and mica surfaces and is compared to experimental observation or discrete particle simulations. To do this, we incorporate excluded volume, short-ranged association, and long-ranged dispersion interactions. Many planar models of electric double layers consider the charged interface to be a hard-wall~\cite{tarazona2008density,roth2016shells}. This produces a sharp-crested density profile for the fluid that does not do well to mimic the structure of physical liquid-solid systems. To represent soft interfaces and the adsorption of molecules to surfaces an external, an integrated form of a Lennard-Jones potential is typically supplied~\cite{siderius2011extension,steele1973physical,balbuena1993theoretical}. We investigate the textured surface of mica by combining an external potential with an explicitly resolved solid density peak, representing the outer solid layer; this permits adsorption to be quantified by a combination of dispersion and site-binding interactions. Upon including ions and electrostatic interactions, the model is able to resolve the formation of the bound Stern layer at high surface charge, charge reversal in the presence of sufficient bulk salt concentration, and ion correlations that lead to strong inter-attractive stresses between like-charged surfaces. Because our model assumes uniform out-of-plane statistics, and integrates free-energy contributions accordingly, the inter-attraction between like-charged surfaces does not arise from in-plane organization of ions.

\section{Methods}

We model the electrolyte that fills the space between negatively charged solid boundaries as a composition of water ($\mathrm{w}$), counterions ($+$), and coions ($-$). Here, we make the choice to include the solvent explicitly to account for the effect that hydration forces play in structuring the electrolyte near the hard or textured interfaces. Our task within cDFT is to define a Helmholtz free energy functional $\mathcal{F} = \mathcal{F}_\mathrm{id}+\mathcal{F}_\mathrm{ex}$, which includes ideal and excess parts, that accurately accounts for the relevant energetic contributions to the electrolyte system. As usual, the ideal part, which accounts for the entropy of mixing of our three-component system, reads
\begin{equation}
\label{eq:free_energy_id}
    \beta \mathcal{F}_\mathrm{id}(\{\rho_i(\boldsymbol{r})\}) = \sum_{i = \mathrm{w,+,-}}\int \mathrm{d}\boldsymbol{r} \left\{\rho_i(\boldsymbol{r})\left[\ln(\Lambda^3\rho_i(\boldsymbol{r}))-1\right]\right\} 
\end{equation}
where $\beta=k_\mathrm{B}T$ sets the thermal energy scale, $\Lambda$ is the de Broglie wavelength, and $\rho_i$ represents the ensemble-averaged number density of species $i$; the factor $\Lambda^3$ is added for dimensional consistency, but does not affect the thermodynamics. The excess part is modeled as a sum that accounts for the repulsive, attractive, and electrostatic interactions in the system and is defined by
\begin{equation}
\mathcal{F}_\mathrm{ex}= \mathcal{F}_\mathrm{hs} +  \mathcal{F}_\mathrm{disp} + \mathcal{F}_\mathrm{assoc} + \mathcal{F}_\mathrm{es}.
\label{eq:excess_free_energy}
\end{equation}

Before providing mathematical expressions of the individual components in Eq.(\ref{eq:excess_free_energy}), we provide a brief background on each. $\mathcal{F}_\mathrm{hs}$ is the free energy due to hard-sphere interactions. It accounts for the finite size of our ions and water molecules and we choose to calculate it using Tarazona's tensorial adaptation of Rosenfeld's Fundamental Measure Theory (FMT) \cite{tarazona2000density, rosenfeld1989free}. Although more accurate modified FMT representations for hard-sphere systems are available, for instance the White Bear formalisms~\cite{roth2010fundamental}, we choose Tarazona's version because it allows for better representation of planar layered systems and crystallization~\cite{tarazona2008density} -- aspects our ongoing works builds on. Moreover, any differences in accuracy are small relative to the assumptions to be made for the other free energy contributions in Eq.(\ref{eq:excess_free_energy}).

The free-energy contributed by long-range, dispersive van der Waals interactions between water molecules, and water molecules and ions is included in $\mathcal{F}_\mathrm{disp}$; inter-ion dispersion interactions are omitted as they are significantly less important than the long-range Coulomb interactions to be described below. The interactions extend across-several particle diameters and are treated and compared under two approaches. In the first approach, a mean-field (MF) approximation explicitly integrates the pair-potential across the density fields of the interacting components. In the second approach, a high-temperature (HT) expansion of the free energy is performed akin to the method of Barker and Henderson~\cite{barker1976liquid,gil1997statistical}. The HT expansion is known to improve the temperature-dependence of the phase diagram, particularly near the critical-point. The method follows the Statistical Associating Fluid Theory of Variable Range (SAFT-VR;~\cite{gil1997statistical}), in which the free energy is expanded about the reference, hard-sphere system and the local compressibility approximation (LCA) is applied to account for fluctuations in the attractive energy that originate from the compressiblity of the system.

Next, $\mathcal{F}_\mathrm{assoc}$ is the free energy due to short-ranged intra-species associative interactions between water molecules and inter-species associative interactions between water molecules and ions. More specifically, the term is used to model the hydrogen-bonding between water molecules and the hydration of ions. As with the dispersive term, we do not permit ions to associate with one another. To measure $\mathcal{F}_\mathrm{assoc}$ we employ an inhomogeneous adaptation of Wertheim's thermodynamic perturbation theory~\cite{wertheim1984fluids_I,wertheim1984fluids_II,wertheim1986fluids_III,wertheim1986fluids_IV}. Specifically, we employ the associating cDFT of Yu and Wu (Yu-Wu)~\cite{yu2002fundamental}, which adequately reproduces Monte Carlo (MC) results of equivalent discrete particle systems~\cite{camacho2020new} and has shown qualitative improvement in representing the density distributions of inter-attractive species relative to the \latin{interfacial} Satistical Associating Fluid Theory (iSAFT) developed by Segura \latin{et al.}~\cite{segura1995associating,segura1997associating}. The two adaptations differ in how the non-local character of site-bonding on the water molecules or ions is treated. Another advantage of the Yu-Wu theory is that it conveniently formulates the association free energy in terms of FMT densities. Proper treatment of $\mathcal{F}_\mathrm{assoc}$ is particularly important in achieving the baseline structure of water adjacent to our material surfaces; although less a focus of the current study, associative interactions also play an important role in delineating the liquid-vapor phase diagram and interfacial tension between gas and liquid phases, which has been successfully applied to a host of other chemicals (See, \textit{e.g.}, Refs.(\citenum{gloor2007prediction,mac2010modeling,pereira2011integrated})).

Lastly, $\mathcal{F}_\mathrm{es}$ includes the electrostatic terms that define the electric potential and incorporate the Coulomb interactions. Within an inhomogeneous cDFT formalism, Coulomb interactions are typically split into i) a contribution that compares the bulk electrostatic free energy to the local electrostatic free energy -- this contribution exists regardless of whether the system is locally electro-neutral -- and ii) a contribution that accounts for long-range interactions that balance the surface charge of the bounding interface and characterizes the electric double layer (EDL) -- which is not locally electro-neutral. To approximate contribution i), success has been found in adapting analytical solutions for the electrostatic free energy of bulk systems under the mean spherical approximation (MSA)~\cite{blum1975mean,blum1987analytical}. For instance, Voukadinova \latin{et al.}~\cite{voukadinova2018assessing} provide a comparison of three adaptations, which they term the bulk fluid (BF)~\cite{kierlik1991density,rosenfeld1993free}, reference fluid density (RFD)~\cite{gillespie2002coupling}, and functionalized MSA (fMSA)~\cite{roth2016shells,jiang2021revisiting} theories. While the first two adaptations build on the idea of performing a density expansion of the electrostatic free energy around a reference system, the latter approach measures the electrostatic free energy directly by considering the ions as spherical capacitors of a radius that depends on a local screening parameter $\Gamma(\boldsymbol{r})$. In the current work, we adopt the fMSA theory as it is computationally more efficient to implement than the RFD theory and more accurate than the BF theory. Lastly, contribution ii) is incorporated under the mean-field assumption by solving the Poisson equation to relate the eletrostatic potential to the charge distribution. Throughout our implementation of $\mathcal{F}_\mathrm{es}$ we consider the medium separating ions as a uniform dielectric continuum, this albeit measuring the interactions with the water molecules in the other free energy terms explicitly. There are ongoing efforts to explicitly represent the mediating effects of water dipoles to the electrostatic energy within cDFT~\cite{de2022polar}.
 
To find the equilibrium density distributions for $\mathrm{w}$, +, and -, the Helmholtz free energy in Eq.\ref{eq:excess_free_energy} is used to measure the grand potential $\Omega$ by way of a Legendre transformation (see, \textit{e.g.}, the textbook of Hansen\cite{hansen2013theory})
\begin{equation}
    \Omega = \mathcal{F}(\{\rho_i(\boldsymbol{r})\}) + \sum_{i = \mathrm{w,+,-}}\int \mathrm{d}\boldsymbol{r} \left\{\rho_i(\boldsymbol{r})\right[V_{\mathrm{ext},i}(\boldsymbol{r}) -\mu_i\left]\right\},
    \label{eq:grand_potential}
\end{equation}
holding temperature, $T$, and the system volume, $V$, constant. In the equation above, $V_{\mathrm{ext},i}$ is an external potential that defines the interactions of the particles with the (mineral) boundaries, and $\mu_i$ is the species-specific chemical potential in a connected, external reservoir. The density profiles are then obtained by using the variational derivative of Eq.(\ref{eq:grand_potential})
\begin{equation}
\frac{\delta \Omega}{\delta \rho_i} = \frac{\delta \mathcal{F}}{\delta \rho_i}+V_{\mathrm{ext},i}-\mu_i=0,
\label{eq:chemical_potential}
\end{equation}
to find the density distributions that equilibrate the chemical potentials within the system $\delta \mathcal{F}/\delta \rho_i + V_\mathrm{ext,i}$ to the chemical potentials outside the system $\mu_i$. Specifically, the equilibrium density profile of each species is related to its bulk density in the reservoir $\rho_{\mathrm{b},i}$ by
\begin{equation}
\rho_i(\boldsymbol{r}) = \rho_{\mathrm{b},i} \exp\left(c_i^{(1)}(\boldsymbol{r})-\beta V_\mathrm{ext,i}(\boldsymbol{r}) +\beta \mu_{\mathrm{ex},i}\right)
\label{eq:equilibrium_densities}
\end{equation}
where $c^{(1)}_i(\boldsymbol{r}) = -\beta (\delta \mathcal{F}_\mathrm{ex}[\{\rho_j(\boldsymbol{r})\}]/\delta \rho_i(\boldsymbol{r}))$ is the one-particle direct correlation function and $\mu_\mathrm{ex,i}$ is the excess part of the chemical potential. Numerical procedures for approximating $\rho_i$ are discussed, for instance, by Roth~\cite{roth2010fundamental} and Härtel\cite{hartel2013density}, and we adopt the method of Picard iteration to solve for the equilibrium profiles throughout this paper.

\subsection{Excess free energy due to hard-sphere repulsion and long-range dispersion interactions}

The shapes of the water molecules and ions are approximated as differently sized spherical particles. We express the pair potential between species, which incorporates steric repulsion and attractive dispersion interactions, as a generalized Mie potential of the form
\begin{equation}
u_{ij}(r) = C\epsilon_{ij}^\mathrm{Mie}\left[\left(\frac{d_{ij}}{r}\right)^{\lambda_1}-\left(\frac{d_{ij}}{r}\right)^{\lambda_2}\right],
\end{equation}
where $\lambda_2<\lambda_1$, $C=[\lambda_1/(\lambda_1-\lambda_2)](\lambda_1/\lambda_2)^{\lambda_2/(\lambda_1-\lambda_2)}$, $r$ is the distance between particle centers, $d_{ij}$ are the \textit{soft} Mie diameters, and $\epsilon_{ij}$ measure the interaction strengths. The potentials have minima at $r_\mathrm{min}^{(i,j)}=(\lambda_1/\lambda_2)^{1/(\lambda_1-\lambda_2)}d_{ij}$. If the radial distribution functions between species, $g_{ij}$, were known, the free energy contribution due to the Mie potential could be calculated by summing the interactions between all species:
\begin{equation}
\mathcal{F}_\mathrm{Mie} = \frac{1}{2}\sum_{i,j=\mathrm{w},+,-}\int \int \mathrm{d}\boldsymbol{r}\mathrm{d}\boldsymbol{r}'\left[\rho_i(\boldsymbol{r})\rho_j(\boldsymbol{r}') g_{ij}(\boldsymbol{r},\boldsymbol{r}')u_{ij}(|\boldsymbol{r}-\boldsymbol{r}'|)\right].
\end{equation}
However, $g_{ij}$ is generally not known and the free energy is instead expanded about the system's hard-sphere reference state~\cite{barker1967perturbation,barker1967perturbationII}. In this way, the hard-sphere reference state provides a template for the structure of the electrolyte and the free energy contributed by the Mie interactions is approximated by
\begin{equation}
\label{eq:HT_expansion}
\mathcal{F}_\mathrm{Mie} \approx \mathcal{F}_\mathrm{hs} + \mathcal{F}_\mathrm{disp}= \mathcal{F}_\mathrm{hs} + \beta \mathcal{F}_1 + \beta^2\mathcal{F}_2 + ..., 
\end{equation}
where we note that $\mathcal{F}_1$ and $\mathcal{F}_2$ have dimensions of (energy)$^2$ and (energy)$^3$, respectively. The higher-order terms are expected to be become less important as the density increases and the fluid becomes nearly incompressible. $\beta \mathcal{F}_1=\mathcal{F}_\mathrm{att}$ is the mean attractive energy, and $\beta^2\mathcal{F}_2=\mathcal{F}_\mathrm{fluc}$ describes the fluctuations of the attractive energy as a consequence of the compression of the fluid.

Next, we decompose the Mie potential $u_{ij}=u_\mathrm{att}^{(i,j)}+u_\mathrm{rep}^{(i,j)}$ into an attractive, perturbation potential, $u_\mathrm{att}^{(i,j)}$, and a repulsive, reference  potential, $u_\mathrm{rep}^{(i,j)}$~\cite{hansen2013theory,barker1976liquid}. The choice of the decomposition depends on whether we treat the attractive potential in a mean-field sense or via a second-order temperature expansion. In the MF case, we choose the Weeks-Chandler-Andersen separation~\cite{weeks1971role}, where the attractive part takes the form $u_\mathrm{att}^{(i,j)}=u_{ij}$ for $r_{ij}>r_\mathrm{min}^{(i,j)}$ and $u_\mathrm{att}^{(i,j)}=-\epsilon_{ij}$ otherwise, and the repulsive part takes the form $u_\mathrm{rep}^{(i,j)}=u_{ij}+\epsilon_{ij}$ for $r_{ij}<r_\mathrm{min}^{(i,j)}$ and $u_\mathrm{rep}^{(i,j)}=0$ otherwise. When opting for the HT-expansion, the method of Barker and Henderson instead separates the perturbation and reference potentials at the change-over in sign of $u_\mathrm{ij}$ at $r=d_{ij}$~\cite{barker1967perturbationII,hansen2013theory}. If the energy and length parameters between water molecules, $\epsilon_\mathrm{ww}$ and $d_\mathrm{ww}$, and between water molecules and cations (anions), $\epsilon_{\mathrm{w+}}$ ($\epsilon_{\mathrm{w-}}$) and $d_\mathrm{w+}$ ($d_\mathrm{w-}$), are known, the temperature-dependent hard-sphere diameters $\sigma_i$ can be approximated by 
\begin{subequations}
\begin{align}
&\sigma_\mathrm{w} = \int \mathrm{d}r\left[\exp(-\beta u_\mathrm{rep}^{^{(\mathrm{w,w})}})-1\right] \\
&\sigma_\mathrm{\pm} = 2\int \mathrm{d}r\left[\exp(-\beta u_\mathrm{rep}^{^{(\mathrm{w,\pm})}})-1\right]-\sigma_\mathrm{w}.
\end{align}
\end{subequations}
Here, the Lorentz combining rule is used to measure the diameters for the ions, which in general are not equal to the diameter of the water molecules, and -- given equal potentials -- the MF and HT hard-sphere diameters differ due the difference in choice of $u_\mathrm{rep}^{(i,j)}$.

The free energy contributions due to hard-sphere repulsion and long-range dispersion interactions are now treated separately.

\subsubsection{Fundamental Measure Theory: Free energy of hard-sphere mixtures}

For hard-sphere fluid mixtures, the celebrated insight by Rosenfeld~\cite{rosenfeld1989free} was to represent particles by a set of fundamental geometric measures that allows the system's excess free-energy density $\Phi_\mathrm{hs}$ to be expressed as a function of weighted densities $n_{\alpha,i}(\boldsymbol{r})$ that reproduce the Percus-Yevick equation-of-state~\cite{percus1958analysis}; an in-depth review of FMT is given by Roth\cite{roth2010fundamental}. In our study, the version of FMT used to calculate $\mathcal{F}_\mathrm{hs}$ is that of Tarazona~\cite{tarazona2000density}, which is summarized by
\begin{equation}
\frac{\mathcal{F}_\mathrm{hs}(\{\rho_i\})}{k_\mathrm{B}T} = \int \mathrm{d}\boldsymbol{r} \left[\Phi_\mathrm{hs}(\{n_\alpha(\boldsymbol{r})\})\right],
\label{eq:hard_sphere_excess}
\end{equation}
where $\Phi_\mathrm{hs} = \Phi_\mathrm{hs}^{(1)}+\Phi_\mathrm{hs}^{(2)}+\Phi_\mathrm{hs}^{(3)}$ and
\begin{subequations}
    \begin{align}
        \Phi_\mathrm{hs}^{(1)} &= -n_0\ln(1-n_3), \\
        \Phi_\mathrm{hs}^{(2)} &= \frac{n_1 n_2 - \boldsymbol{n}_1 \cdot \boldsymbol{n}_2}{1-n_3}, \\
        \Phi_\mathrm{hs}^{(3)} &= \frac{1}{24\pi(1-n_3)^2}\left[n_2^3 - 3 n_2 \boldsymbol{n}_2\cdot\boldsymbol{n}_2 + \frac{9}{2}\left(\boldsymbol{n_2}\cdot \mathcal{T}_2\cdot \boldsymbol{n}_2 - \frac{1}{2}\mathrm{Tr}(\mathcal{T}_2^3)\right)\right].
    \end{align}
\end{subequations}
The equations above include four weighted scalar densities $n_\alpha$ with $\alpha \in \{0,1,2,3\}$, two vector densities $\boldsymbol{n}_\alpha$ with $\alpha\in\{1,2\}$, and one tensor density $\mathcal{T}_2$; in the remainder of the paper the full set of densities is collectively represented by $\{n_\alpha\}$. Each of these \textit{global} or \textit{summary} densities is calculated as the composite of the densities of the individual species $n_\alpha=\sum_{i=\mathrm{w},+,-}n_{\alpha,i}$, where, for spherical particles,
\begin{equation}
\label{eq:weighted_densities}
n_{\alpha,i}(\boldsymbol{r}) = \int \mathrm{d}\boldsymbol{r}'\left[\omega_{\alpha,i}(\boldsymbol{r}-\boldsymbol{r}')\rho_i(\boldsymbol{r}')\right],
\end{equation}
and the four independent weight functions $\{\omega_{{\alpha},i}\}$ pertaining to $\{n_{\alpha,i}\}$ read as
\begin{subequations}
    \begin{align}
        &\omega_{3,i}(r) = \Theta(\sigma_i-r),\\
        &\omega_{2,i}(r) = \delta(\sigma_i-r),\\
        &\boldsymbol{\omega}_{2,i}(\boldsymbol{r}) = \frac{\boldsymbol{r}}{r}\delta(\sigma_i-r),\\
        &\mathcal{W}_{2,i}(\boldsymbol{r})= \left(\frac{\boldsymbol{r}\otimes\boldsymbol{r}}{r^2}-\frac{1}{3}\mathbf{1}\right)\omega_{2,i}(r).
    \end{align}
\end{subequations}
The remaining weight functions are readily calculated from the former as follows: $\omega_{1,i}(r)=\omega_{2,i}/(4\pi\sigma_i)$, $\omega_{0,i}(r)=\omega_{2,i}(r)/(4\pi\sigma_i^2)$, and $\boldsymbol{\omega}_{1,i}(r) = \boldsymbol{\omega}_{2,i}(r)/(4\pi\sigma_i)$. Above, $\Theta(r)$ denotes the Heaviside step function, $\delta(r)$ is the Dirac-delta distribution, $r=|\boldsymbol{r}|$ denotes the distance from the center of a particle placed at the origin, and $\sigma_i$ is the diameter of species $i$. The bolded weight functions relate to the vector weighted densities, and calligraphic font has been used to signify the tensorial weight function. Importantly, integrating $\omega_{\alpha,i}$ over space gives the volume $V_i$ ($\alpha=3$), surface area $S_i$ ($\alpha=2$), radius $a_i$ ($\alpha=1$), and Euler characteristic ($\alpha=0$) of species $i$. Accordingly, integrating the weight functions over the homogeneous bulk densities $\rho_{\mathrm{b},i}$ of the species in the reservoir provides $n_{3,i}^\mathrm{b}=(4\pi a_i^3/3)\rho_{\mathrm{b},i}$, $n_{2,i}^\mathrm{b}=(4\pi a_i^2)\rho_{\mathrm{b},i}$, $n_{1,i}^\mathrm{b}= a_i \rho_{\mathrm{b},i}$, and $n_{0,i}^\mathrm{b}=\rho_{\mathrm{b},i}$. These bulk densities are typically referred to as the scaled particle theory (SPT) variables, for which $n_3^\mathrm{b}=\sum_{i}n_{3,i}^\mathrm{b}$ measures the bulk packing density $\eta^\mathrm{b}$ of the fluid. Calculation of the weighted densities using Eq.(\ref{eq:weighted_densities}) is performed efficiently using Fourier transforms as seen, for instance, in the appendix of the dissertation of Härtel~\cite{hartel2013density}.

\subsubsection{Barker-Henderson expansion: Free energy pertaining to long-range dispersion interactions}

The first-order term pertaining to the long-range dispersion interactions in the high-termperature expansion presented in Eq.(\ref{eq:HT_expansion}) reads according to Barker and Henderson as follows~\cite{gil1997statistical,barker1967perturbationII}
\begin{equation}
\label{eq:F_att}
\mathcal{F}_\mathrm{att} = \frac{1}{2}\sum_{i,j=\mathrm{w},+,-}\int \mathrm{d}\boldsymbol{r}  \int \mathrm{d}\boldsymbol{r}'\left[ \rho_i(\boldsymbol{r}) \rho_\mathrm{j}(\boldsymbol{r}') g^{(i,j)}_\mathrm{ref}(\boldsymbol{r},\boldsymbol{r}';\{\rho_\alpha(\boldsymbol{r}),\rho_\beta(\boldsymbol{r}')\}) u_\mathrm{att}(|\boldsymbol{r}-\boldsymbol{r}'|)\right],
\end{equation}
where the radial distribution function is typically approximated by that of the reference system of our hard-sphere mixture, $g_\mathrm{ref}=g_\mathrm{hs}$. An expression for $g_\mathrm{ref}$ under system-specific conditions is generally not known \textit{a priori} and thus further simplification is needed. In the mean-field (MF) approximation, we set $g_\mathrm{ref}=1$ and consequently also neglect the second-order term, $\beta^2\mathcal{F}_2$. These assumptions correspond to dismissing any correlations between density fields. 

In a second approach, following the steps outlined by the SAFT-VR model~\cite{gil1997statistical,gross2001perturbed,llovell2010classical}, the attractive potential is split into two Sutherland potentials, $\psi_{ij}(r=|\boldsymbol{r}-\boldsymbol{r}'|,\lambda)=(d_{ij}/r)^\lambda$, such that
\begin{equation}
\label{eq:g_hs_u_att}
g^{(i,j)}_\mathrm{ref}u_\mathrm{att}=
\begin{cases}
    C\epsilon^\mathrm{Mie}_{ij} g^{(i,j)}_\mathrm{hs}(r;\{\eta(\boldsymbol{r}),\eta(\boldsymbol{r}')\})\left[\psi_{ij}(r,\lambda_1) + \psi_{ij}(r,\lambda_2)\right] & \text{for } r>d_{ij} \\
    0 & \text{for } r\le d_{ij}
\end{cases}.
\end{equation}
While the hard-sphere pair correlation function has known, accurate representations for homogeneous systems, a functional representation is difficult to obtain for inhomogeneous systems. Thus, we choose to evaluate $g^{(i,j)}_\mathrm{hs}(r,\bar{\eta})$ using the arithmetic mean of the packing densities at the two points of evaluation, $\boldsymbol{r}$ and $\boldsymbol{r}'$: $\bar{\eta}(\boldsymbol{r},\boldsymbol{r}')=[n_3(\boldsymbol{r})+n_3(\boldsymbol{r}')]/2$. With this simplification, the expression in Eq.(\ref{eq:g_hs_u_att}) is substituted into Eq.(\ref{eq:F_att}), for which the inner integral for each Sutherland potential is written as
\begin{equation}
\label{eq:att_integral}
C\epsilon^\mathrm{Mie}_{ij} \rho_i(\boldsymbol{r}) \int_{r \ge d_{ij}} \mathrm{d}\boldsymbol{r}'\left[ g^{(i,j)}_\mathrm{hs}(r,\bar{\eta}(\boldsymbol{r},\boldsymbol{r}'))\rho_{j}(\boldsymbol{r}')  \psi_{ij}(r,\lambda)\right].
\end{equation}
Ordinarily, for homogeneous fluids, $g_\mathrm{hs}^{(i,j)}$ can be pulled out of the integral by use of the mean-value theorem. However, the choice of $\bar{\eta}$ requires the packing density to be evaluated using the packing densities at $\boldsymbol{r}$ and $\boldsymbol{r}'$. As was done by Llovell \latin{et al.}~\cite{llovell2010classical}, we apply the homogeneous parameterization of the radial distribution function of Gil-Villegas \latin{et al.}~\cite{gil1997statistical} to our inhomogeneous system. The parameterization proceeds as follows: The pair correlation function assumes the shape of its contact value
\begin{equation}
\label{eq:mean_value_approx}
g_\mathrm{hs}(\xi,\bar{\eta}(\boldsymbol{r},\boldsymbol{r}')) = g_\mathrm{hs}(1,\bar{\eta}^\mathrm{eff}(\boldsymbol{r},\boldsymbol{r}',\lambda)), 
\end{equation}
evaluated at $\bar{\eta}^\mathrm{eff} < \bar{\eta}$. Above, $\xi$ corresponds to the ``mean-value" inter-particle distance at packing fraction $\bar{\eta}$; in other words, for uniform densities $\rho_i$ and $\rho_j$, $g_\mathrm{hs}(\xi,\bar{\eta})\int \mathrm{d}\boldsymbol{r}(\psi_{ij}) = \int \mathrm{d}\boldsymbol{r}(g_\mathrm{hs}(r,\bar{\eta})\psi_{ij})$. The relation in Eq.(\ref{eq:mean_value_approx}) is justified by the observation that $g_\mathrm{hs}$ is a monotonically decreasing function of $\xi$, which implies that a functional relation of $g_\mathrm{hs}(\xi,\bar{\eta}(\boldsymbol{r},\boldsymbol{r}'))$ in terms of $\bar{\eta}^\mathrm{eff}$ is possible. In fact, this relationship was established in Ref.(\citenum{gil1997statistical}) using Carnahan-Starling's analytic expression for the contact value~\cite{carnahan1969equation}, 
\begin{equation}
\label{eq:hs_contact_correlation}
g_\mathrm{hs}(1,\bar{\eta}^\mathrm{eff}) = \frac{1-\bar{\eta}^\mathrm{eff}/2}{(1-\bar{\eta}^\mathrm{eff})^2},
\end{equation}
with an effective density that is selected as a quadratic function of the true density
\begin{equation}
\label{eq:effective_packing}
\bar{\eta}^\mathrm{eff}(\lambda) = c_1(\lambda)\bar{\eta} + c_2(\lambda)\bar{\eta}^2.
\end{equation}
Above, the coefficients $c_1$ and $c_2$, are calculated by solving Eq.(\ref{eq:att_integral}) for homogeneous systems and are listed in the Supporting Information (SI)\footnote{
Certainly, it is also possible to calculate Eq.(\ref{eq:att_integral}) by substituting an approximation for the radially-dependent $g_\mathrm{hs}(r,\bar{\eta})$, though these expressions are complicated and do not simplify the calculation.
}.

The second-order fluctuation term, is derived from the \textit{local} compressibility approximation~\cite{barker1967perturbation,gil1997statistical}

\begin{equation}
\label{eq:F_fluc}
\begin{split}
\mathcal{F}_\mathrm{fluc} &= -\frac{\beta}{4}\sum_{i,j=\mathrm{w},+,-}\int \mathrm{d}\boldsymbol{r} \left\{\rho_i(\boldsymbol{r}) \rho_j(\boldsymbol{r}) \int \mathrm{d}\boldsymbol{r}'\left[u_\mathrm{att}(|\boldsymbol{r}-\boldsymbol{r}'|)^2 k_\mathrm{B}T\left(\frac{\partial (\eta(\boldsymbol{r}) g^{(i,j)}_\mathrm{ref}(\boldsymbol{r}',\eta)}{\partial P_\mathrm{ref}}\right)\right]\right\} \\
&= -\frac{\beta}{4}\sum_{i,j=\mathrm{w},+,-}\int \mathrm{d}\boldsymbol{r} \left\{K_\mathrm{hs,0}(\boldsymbol{r},\eta) \rho_i(\boldsymbol{r}) \rho_j(\boldsymbol{r}) \frac{\partial}{\partial \eta} \left[\eta \int \mathrm{d}\boldsymbol{r}'\left(g^{(i,j)}_\mathrm{ref}(\boldsymbol{r}',\eta)u_\mathrm{att}(|\boldsymbol{r}-\boldsymbol{r}'|)^2 \right)\right]\right\} 
\end{split}
\end{equation}
where $K_\mathrm{hs,0}=k_\mathrm{B}T \partial \eta/\partial P_\mathrm{hs}$ measures the local compressibility of the reference system, again approximated with the monodisperse hard-sphere result,
\begin{equation}
K_\mathrm{hs,0} = \frac{(1-\eta)^4}{(1+2\eta)^2} =\frac{(1-n_3)^4}{(1+2n_3)^2}  .
\end{equation}
The inner integral in Eq.(\ref{eq:F_fluc}) is further simplified by assuming a spherically symmetric density profile and that the reference correlation function takes on the contact value expressed in Eq.(\ref{eq:hs_contact_correlation}). These simplifications lead to

\begin{equation}
\begin{split}
\int \mathrm{d}\boldsymbol{r}'\left(g_\mathrm{hs}^{(i,j)}(\boldsymbol{r}',\{\varsigma_\alpha\})u_\mathrm{att}(|\boldsymbol{r}-\boldsymbol{r}'|)^2 \right) &\approx 4\pi r_\mathrm{min}^{(i,j)3} C^2 \epsilon_{ij}^2 g_\mathrm{hs}^{(i,j)}(\xi) \int_{1}^\infty \mathrm{d}\hat{r}\left[\hat{r}^2  \psi^2(\lambda_2)\right] \\
&= 4\pi r_\mathrm{min}^{(i,j)3} g_\mathrm{hs,0}(\eta^\mathrm{eff}) \frac{C^2 \epsilon_{ij}^2}{\lambda_2-3}
\end{split}
\end{equation}
so long as $\lambda_2>3$, and the term corresponding to the repulsive Sutherland potential is assumed to be  comparatively small in the domain $\hat{r}=r/r_\mathrm{min}>1$. Thus, the final form of the fluctuation term implemented for the dispersion interactions reads as
\begin{equation}
\mathcal{F}_\mathrm{fluc} = -\frac{\pi\epsilon_{ij}^2K_\mathrm{hs,0} C^2 r_\mathrm{min}^{(i,j)3}}{k_\mathrm{B}T(\lambda_2-3)} \sum_{ij}\int \mathrm{d}\boldsymbol{r}\left[\rho_i(\boldsymbol{r})\rho_j(\boldsymbol{r})\left( g_\mathrm{hs,0}^{(i,j)} + n_3 \frac{\partial g_\mathrm{hs,0}^{(i,j)}}{\partial n_3}\right)\right].
\end{equation}

Above, we chose not to use the contact correlation function for hard-sphere mixtures~\cite{boublik1970hard,mansoori1971equilibrium} (though the mixture expression is adopted below when calculating the free energy relating of the associative interactions), as it invokes evaluating many additional convolution integrals for $\beta\delta \mathcal{F}_1/\delta \rho_i$ and $\beta^2\delta \mathcal{F}_2/\delta \rho_i$ at each iteration, while minimally modifying the predicted density distributions\footnote{We implemented contact correlation functions for monodisperse and polydisperse hard-sphere systems and found little change.}. Additionally, it is worth noting that the effective packing density $\eta^\mathrm{eff}$ used in $g_\mathrm{hs,0}$ is calculated assuming hard-spheres of diameter $\sigma_{ij}$, while particle contact is initiated at $d_{ij}$ in Eq.(\ref{eq:F_att}) following the threshold set by Eq.(\ref{eq:g_hs_u_att}).

\subsection{Excess free-energy due to particle association}

Thus far, our system corresponds to an inhomogeneous van der Waals fluid that accounts for steric repulsion and long-range dispersion interactions. The development of the electrolyte model is continued by introducing short-range associative interactions. \textit{Short-range}, here, implies interactions that are confined to a particle's nearest neighbors and \textit{associative} refers to interactions that are directed, dependent on particle orientation, and are thus able to initiate coordination complexes between atoms and molecules. For instance, in the case of hydrogen bonding between H$_2$O molecules, the partial positive charges on the hydrogen atoms temporarily \textit{bond} to the lone electron pairs on the oxygen atom. Thus, long-range, many-body, instantaneous correlations between water molecules can form aggregates of water molecules that influence the dielectric properties of the solvent~\cite{kirkwood1939dielectric,bonthuis2011dielectric}. Ongoing efforts -- not presented here -- are geared toward explicitly connecting associative interactions to these dipole correlations. In fact, it is worth mentioning that bulk associative MSA models that explicitly incorporate the effects of the solvent dipoles into the electrostatic free energy exist~\cite{blum1978solution,blum1987analytical,simonin2020solution,simonin2021full, holovko2018application}, though, to our knowledge, they have yet to be extended to interfacial systems.

For the model chosen in this study, association between species is not rigidly coupled to orientation, as the orientation field, in general, is not resolved. Instead, according to Wertheim's TPT~\cite{wertheim1984fluids_I,wertheim1984fluids_II,wertheim1986fluids_III,wertheim1986fluids_IV}, each species is assigned sites (or patches) that are uniformly distributed around the particles' surfaces and restricted to interacting with sites found on other ions and molecules. Further, each site covers only a small portion of the particle's surface, defined by an opening angle $\theta_\mathrm{c}$, and extends away from the surface of the particle by a fraction $\delta$ of its radius. Bonds between sites $\alpha$ on particles $i$ and sites $\beta$ on particles $j$ are formed via square-well potentials of magnitude $\epsilon_\mathrm{\alpha\cdot\beta}^{(i,j)}$. For the case of water, we implement the so-called 4C model~\cite{huang1990equation}, which supposes that each molecule has four association sites of two distinctive types: Two sites designating the lone pair of electrons, indicated by $e_0$, on the oxygen atom, and two sites of partial positive charge on the hydrogen atoms $\mathrm{H}$. Only sites of opposite type are able to bond. We summarize this bonding behavior by $\epsilon_{e_0 \cdot e_0} = \epsilon_\mathrm{H \cdot H} = 0$ and $\epsilon_{e_0\cdot \mathrm{H}} > 0$. Lastly, we also allow the ions in the electrolyte to bond to water molecules according to their charge. These additional non-zero, associative interaction energies are designated by $\epsilon_{+\cdot e_0}$ and $\epsilon_{-\cdot \mathrm{H}}$. According to these choices, interacting sites have no fixed position relative to the particle orientations and their bonding instead depends on the distribution of particle surfaces within a site's reach.

\subsubsection{Yu and Wu's adaptation of Wertheim's TPT for associating fluids}

We implement the association model of Yu and Wu~\cite{yu2002fundamental} as it performed better than other inhomogeneous cDFT association models we trialed. In their derivation, the SPT variables in Wertheim's expression of the Helmholtz free energy for bond formation are replaced by the equivalent FMT weighted densities. The resulting expression is
\begin{equation}
\beta \mathcal{F}_\mathrm{assoc}^\mathrm{Y-W} = \int \mathrm{d}\boldsymbol{r}\left\{\sum_{i=\mathrm{w},+,-} n_{2,i} \zeta_i\sum_{\alpha\in \Gamma_i} \left[ \ln(\chi_\alpha^{(i)}) + \frac{1-\chi_\alpha^{(i)}}{2} \right]\right\},
\label{eq:assoc_YuWu}
\end{equation}
where the first term in the brackets accounts for the entropy of unbonded sites and the second term accounts for the energy change imbued by bond formation. Above, $\chi_\alpha^{(i)}(\boldsymbol{r})$ represents the fraction of particles of species $i$ located at position $\boldsymbol{r}$ not bonded at sites of type $\alpha$, and $\zeta_i=1-\mathbf{n}_{2,i}\cdot\mathbf{n}_{2,i} / n_{2,i}^2$ is a phenomenological correction factor to the particle density and site localization that improves the match between cDFT theory and discrete particle simulations. We note that Stopper \latin{et al.}~\cite{stopper2018bulk} -- who investigated association between patchy colloidal particles -- proposed a modification to the correction factor, replacing $\zeta_i$ with $\zeta_i^3$, which showed improvement in resolving the radial distribution around spherical test particles; we implemented this approach for our planar density profiles and found better comparisons to the simulation data of water when sticking to the original formulation by Yu and Wu. 

Site bonding between the species in the electrolyte is summarized by the following set of mass-action laws:
\begin{subequations}
\label{eq:mass_action_YuWu}
\begin{align}
    &\chi_{e_0}^\mathrm{w} + 2n_{0,\mathrm{w}}\zeta_\mathrm{w} \chi_{e_0}^\mathrm{w}\chi_{\mathrm{H}}^\mathrm{w} \Delta_{e_0\cdot \mathrm{H}} + M_\mathrm{+} n_{0,\mathrm{+}}\zeta_\mathrm{+} \chi_{e_0}^\mathrm{w}\chi^{\mathrm{+}} \Delta_{e_0\cdot \mathrm{+}} = 1\\
    &\chi_\mathrm{H}^\mathrm{w} + 2n_{0,\mathrm{w}}\zeta_\mathrm{w} \chi_{e_0}^\mathrm{w}\chi_{\mathrm{H}}^\mathrm{w} \Delta_{e_0\cdot \mathrm{H}} + M_\mathrm{-} n_{0,\mathrm{-}}\zeta_\mathrm{-} \chi_\mathrm{H}^\mathrm{w}\chi^{\mathrm{-}} \Delta_{\mathrm{H}\cdot \mathrm{-}} = 1\\
    & \chi^{+} + M_+ n_{0,\mathrm{w}}\zeta_\mathrm{w} \chi_{e_0}^\mathrm{w}\chi^\mathrm{+} \Delta_{e_0\cdot \mathrm{+}} = 1\\
    & \chi^{-} + M_- n_{0,\mathrm{w}}\zeta_\mathrm{w} \chi_\mathrm{H}^\mathrm{w}\chi^\mathrm{-} \Delta_{\mathrm{H}\cdot \mathrm{-}} = 1.
\end{align}
\end{subequations}
In these equations, $M_{+/-}$ counts the number of bonding sites on the cations/anions and $\Delta_{\alpha \cdot \beta}$ are interaction terms that depend on the bond volume and bond strength. As per the definition of a mass-action law, the additive terms in Eqn.(\ref{eq:mass_action_YuWu}) represent either a fraction of unbonded sites or the fraction of bonds between sites of a particular type that collectively sum to 1. The site bonding interaction term is approximated by integrating the pair direct correlation function between two particles over the bond volume to arrive at~\cite{segura1995associating} 
\begin{equation}
\Delta^\mathrm{Y-W}_{\alpha \cdot \beta}(\{n_\alpha(\boldsymbol{r})\})= K_{ij}g_\mathrm{hs}^{(i,j)}(\{n_\alpha(\boldsymbol{r})\})\left[\exp\left(\beta \epsilon_{\alpha \cdot \beta}^{(i,j)}\right)-1\right].
\end{equation} 
Here, $K_{ij}=(4\pi/3)(R_i + R_j)^3 \sin(\theta_\mathrm{c}/2)^4[(1+\delta)^3-1]$ measures the bond volume, for which we choose $\sin(\theta_\mathrm{c}/2)=0.229$ and $\delta = 0.05$ unless otherwise noted, and the pair correlation value $g_\mathrm{hs}^{(i,j)}$ is estimated to be that of the hard-sphere mixture at contact~\cite{mansoori1971equilibrium,boublik1970hard,yu2002fundamental,reed1973applied},
\begin{equation}
g_\mathrm{hs}^{(i,j)} = \frac{1}{1-n_3} + \frac{\sigma_i\sigma_j}{2(\sigma_i+\sigma_j)}\frac{n_2 \zeta}{(1-n_3)^2} + \frac{1}{18}\left(\frac{\sigma_i\sigma_j}{\sigma_i+\sigma_j}\right)^2\frac{n_2^2 \zeta}{(1-n_3)^3}.
\end{equation}
Lastly, $\exp(\beta \epsilon_{\alpha\cdot \beta}^{(i,j)})-1$ is the Mayer $f$-function as calculated from the square-well interaction potential.

\subsection{Excess free-energy due to electrostatic interactions}

Roth and Gillespie recently proposed a model that accurately reproduces MC simulation results of multivalent co- and counterion profiles near charged hard walls ~\cite{roth2016shells}. The authors adapted the analytic solution of the MSA for bulk electrolytes~\cite{blum1975mean,blum1987analytical}, to inhomogeneous case of ions in a continuum dielectric. Their central innovation was to treat ions as shells of charge -- the charges of ions were smeared over spheres with a radius depending on the screening length of the electrolyte. We investigate this model in the context of an explicitly resolved solvent near the soft graphene and mica interfaces described above. Importantly, as was done by Roth and Gillespie\cite{roth2016shells}, the dielectric properties of water enter the electrostatic theory as a uniform continuum via the relative electric permittivity, $\varepsilon_\mathrm{r}\approx 78.4$. This is a simplification, as the structure of water dipoles in the vicinity of interfaces is known to create an oscillatory and anisotropic permittivity field~\cite{bonthuis2011dielectric}. For a thorough account of the adopted electrostatic theory, we encourage the reader to explore the original article by Roth and Gillespie\cite{roth2016shells} and a follow-up study that corrects for the theoretical treatment of fully overlapping charged spheres~\cite{jiang2021revisiting}. Nonetheless, we provide a brief recount.

The electrostatic component of the excess free energy in Eq.(\ref{eq:excess_free_energy}) is approximated by
\begin{equation}
\label{eq:electrostatic_free_energy}
\begin{split}
\mathcal{F}_\mathrm{es} = \int \mathrm{d}\boldsymbol{r}\Bigg[&\Phi_\mathrm{es}(\{q_{i}(\boldsymbol{r})\})
+ \frac{1}{2}\sum_{ij} \int \mathrm{d}\boldsymbol{r'}\left\{\rho_i(\boldsymbol{r})\rho_j(\boldsymbol{r}')\Delta \psi_{ij}(|\boldsymbol{r}-\boldsymbol{r}'|)\right\}\\
+& \sum_i z_i e_0 \rho_i(\boldsymbol{r}) \phi(\boldsymbol{r}) \Bigg].
\end{split}
\end{equation}
The first term in the integrand is an inhomogeneous electrostatic energy density that is adapted to the analytic MSA solution~\cite{blum1975mean,blum1987analytical} within the framework of FMT, expressed in the following way:
\begin{equation}
\label{eq:electrostatic_energy_density}
\begin{split}
\Phi_\mathrm{es}(\{q_{i}(\boldsymbol{r})\}) &= \frac{\Gamma(\{q_{i}(\boldsymbol{r})\})^3}{3 \pi} \\
&- \lambda_\mathrm{B} \sum_{i=+,-}  \left(\frac{\Gamma(\{q_{i}(\boldsymbol{r})\}) z_i^2 q_{i}(\boldsymbol{r}) + \eta(\{q_{i}(\boldsymbol{r})\}) z_i q_{i}(\boldsymbol{r})}{1+\Gamma(\{q_{i}(\boldsymbol{r})\}) \sigma_i}\right).
\end{split}
\end{equation}
Above, $\lambda_\mathrm{B}=e_0^2/(4\pi \varepsilon_\mathrm{r}\varepsilon_0 k_\mathrm{B}T)$ is the Bjerrum length with $\varepsilon_0$ the permittivity of free space, $e_0$ is the electron charge, $z_i$ are the ion valances,  $\Gamma$ is the MSA screening parameter, and $q_{i}$ are FMT-like weighted densities that locate the smeared charge of the ions,
\begin{equation}
\label{eq:smeared_charge}
q_{i}(\boldsymbol{r}) = \int \mathrm{d}\boldsymbol{r}\left(\rho_i(\boldsymbol{r}')\frac{\delta(|\boldsymbol{r}-\boldsymbol{r}'|-b_i)}{4\pi b_i} \right),
\end{equation}
where $b_i=(\sigma_i + 1/\Gamma_\mathrm{ref})/2$ are the radii of the charged shells. As was done by Roth and Gillespie, we choose the screening parameter $\Gamma_\mathrm{ref}$ to correspond to that of the bulk fluid when evaluating the convolution integrals, though we need to make adjustments when modeling the electrolytes in confined geometries. Choosing constant shell radii, $b_i$, in Eq.(\ref{eq:smeared_charge}) greatly reduces the computational cost: The convolution window-size is not a function of space and does not need to be adjusted from one iteration to the next. Nonetheless, the nonlocal, implicit expression for $\Gamma$ can be used when calculating $\Phi_\mathrm{es}$ in Eq.(\ref{eq:electrostatic_energy_density}):
\begin{equation}
\Gamma(\{q_{i}(\boldsymbol{r})\}) = \pi \lambda_\mathrm{B}\sum_i \rho_i\left(\frac{z_i - \eta(\{q_i(\boldsymbol{r})\})\sigma_i^2}{1+\Gamma(\{q_i(\boldsymbol{r})\}) \sigma_i}\right)^2.
\end{equation}
Herein, the parameter $\eta$ takes the following form:
\begin{equation}
\eta(\{q_i(\boldsymbol{r})\}) = \frac{1}{H(\{q_i(\boldsymbol{r})\})} \sum_i \frac{z_i q_i(\boldsymbol{r}) \sigma_i}{1+\Gamma(\{q_{i}(\boldsymbol{r})\}) \sigma_i}
\end{equation}
with
\begin{equation}
H(\{q_i(\boldsymbol{r})\}) = \sum_i \frac{q_i(\boldsymbol{r}) \sigma_i^3}{1+\Gamma(\{q_{i}(\boldsymbol{r})\}) \sigma_i} + \frac{2}{\pi}\left(1-\frac{\pi}{6}\sum_i q_i(\boldsymbol{r}) \sigma_i^3 \right).
\end{equation}
For the restricted primitive case (RPM) -- that is, for electrolytes that contain equisized ions ($\sigma_i=\sigma$ for $i=+,-$) of equal charge magnitude ($|z_+|=|z_-|$) -- this parameter vanishes ($\eta=0$). If the electrolyte is also homogeneous, Eq.(\ref{eq:electrostatic_energy_density}) can be interpreted as measuring the energy of spherical capacitors of radius $b_i=(\sigma_i + 1/\Gamma(\{\rho_i\}))/2$.

In inhomogeneous conditions, the long-ranged Coulombic interactions between the charged shells needs to be assessed; such a contribution vanishes under homogeneous conditions due to the symmetry of the energetic contributions between anions and cations under charge-neutrality. We note that within the MSA-closure the interactions between smeared charges assume the form $\psi^\mathrm{sh}_{ij} = z_i z_j e_0^2/(16\pi \varepsilon_\mathrm{r}\varepsilon_0)\times (2r(b_i+b_j)-r^2-(b_i-b_j)^2)/(b_i b_j r)$ when the hard-cores overlap ($r\le\sigma_{ij}$), while the interactions assume the standard Coulomb form $\psi_{ij}^\mathrm{C}=z_i z_j e_0^2 / (4\pi \varepsilon_\mathrm{r}\varepsilon_0 r)$ when the cores do not overlap ($r>\sigma_{ij}$)~\footnote{There is inconsistency in the formulation of the MSA electrostatic energy: The internal energy is calculated assuming the standard Coulomb potential begins when charged shells stop overlapping, while the MSA closure assumes the Coulomb potential begins at the hard-core cutoff~\cite{roth2016shells}.}. In doing so, the electrostatic potential, $\phi$, can be calculated from the Poisson equation under mean-field assumptions,
\begin{equation}
-\varepsilon_\mathrm{r}\varepsilon_0 \nabla^2 \phi = e_0\sum_i z_i \rho_i(\boldsymbol{r}) + f_\mathrm{b}(\boldsymbol{r}),
\end{equation}
where $f_\mathrm{b}(\boldsymbol{r})$ specifies the bound charge along the solid boundaries. Subsequently, Coulomb interactions must be subtracted in the region for which the ions' hard-cores overlap, while the interactions pertaining to the overlapping charged shells are added. This is taken into account in the second term on the right-hand-side of Eq.(\ref{eq:electrostatic_free_energy}), where the expression for $\Delta\psi_{ij}=\psi^\mathrm{sh}_{ij}-\psi^\mathrm{C}_{ij}$ is given by
\begin{equation}
\beta \Delta \psi_{ij}(|\boldsymbol{r}-\boldsymbol{r}'|) = 
\begin{cases}
\displaystyle
-\frac{z_i z_j \bar{\lambda}_\mathrm{B}}{4b_ib_j} \frac{(|\boldsymbol{r}-\boldsymbol{r}'|-b_{ij})^2}{|\boldsymbol{r}-\boldsymbol{r}'|}&\text{if} \quad \Delta b_{ij} \le |\boldsymbol{r}-\boldsymbol{r}'| < \sigma_{ij}\\
\displaystyle
z_iz_j\bar{\lambda}_\mathrm{B}\left(\frac{1}{\max\{b_i,b_j\}}-\frac{1}{|\boldsymbol{r}-\boldsymbol{r}'|}\right)&\text{if} \quad |\boldsymbol{r}-\boldsymbol{r}'| < \Delta b_{ij},
\end{cases}
\end{equation}
which includes the theoretical correction for the case in which one spherical shell sits entirely inside another\cite{jiang2021revisiting}, and $\Delta b_{ij}=|b_i-b_j|$.
In a planar geometry, after integrating over the $y-$ and $z-$ directions, the above potential evaluates to:
\begin{equation}
\begin{split}
&\beta \Delta \psi_{ij}(r)  \\ 
&=\begin{cases}
\displaystyle
\frac{z_i z_j \pi \bar{\lambda}_\mathrm{B}}{6b_i b_j} \left[(r-b_{ij})^3+(b_{ij}-\sigma_{ij})^3\right]& \text{if} \quad |\Delta b_{ij}| < r < \sigma_{ij}\\
\displaystyle
\frac{z_i z_j \pi \bar{\lambda}_\mathrm{B}}{6b_i b_j} \left[(r-b_{ij})^3+(b_{ij}-\sigma_{ij})^3-(r-|\Delta b_{ij}|)^3\right] & \text{if} \quad r < |\Delta b_{ij}|
\end{cases}
\end{split}
\end{equation}
with $r=|x-x'|$.

\section{Results and Discussion}

Equilibrium 1D density profiles for monodisperse and bidisperse hard-sphere systems calculated using Eq.\ref{eq:hard_sphere_excess} for the excess free-energy $\mathcal{F}_\mathrm{ex}$ in Eq.\ref{eq:chemical_potential}-\ref{eq:equilibrium_densities} are presented in Figure S1(a) and Figure S1(b), respectively. The figure panels show density profiles near a hard wall, where $V_{\mathrm{ext},i}=\infty$ for $x<=(\sigma_i/2)$ and $V_{\mathrm{ext},i}=0$ otherwise. For the bidisperse system, a 3:1 size ratio between large and small particles was prescribed to demonstrate FMT's ability to accurately reproduce densities for geometric parameter ranges that exceed the ratios between species in typical geochemical systems; for a MgCl$_2$ salt system, for instance, $\sigma_\mathrm{Mg^{2+}}\approx 1.72$ \r{A} and $\sigma_\mathrm{Cl^-}\approx 3.34$ \r{A}, which provides close to a 1:2 ratio. In the figure, the theoretical cDFT model is compared to MC simulation data of equivalent discrete particle systems. It is readily observed that theory and simulation agree excellently, independent of both the bulk density and size disparity between species. When considering the comparison, note that for bulk water with a concentration of $\rho_\mathrm{b,\mathrm{w}}=56\,\mathrm{M}$ (a value representative of its fluid phase at standard temperature and pressure) and size $\sigma_\mathrm{w}=3$ \r{A}, the normalized density is approximately $\rho_\mathrm{b,w}\sigma_\mathrm{w}^3=0.91$.

To evaluate the performance of the Yu-Wu model, we studied a one-component fluid with a 4C association scheme in contact with a hard-wall and compared it to the density profiles provided by the iSAFT model (see the Supporting Information (SI) for details); for now, fluid molecules are assumed not to associate with the wall. In evaluating the excess chemical potential due to particle association, use is made of the mathematical simplifications provided by Michelsen and Hendriks~\cite{michelsen2001physical} (see SI for details). Figure(S2) plots the predictions of both cDFT models for fluids of (a) moderate and (b) high association strength, and compares the results to discrete particle simulations. As anticipated, the structure of the fluid is dictated primarily by hard-sphere interactions; it takes significant association energy, particularly at densities relevant to aqueous suspensions ($\rho_\mathrm{b,w}\sigma_\mathrm{w}^3 \gtrsim 0.9$), to subdue the density peaks. For dense fluids, significant qualitative differences in the overall structure of the fluid become readily apparent once $\epsilon_{e_0 \cdot \mathrm{H}}\gtrsim 5 \, k_\mathrm{B}T$, for which, in general, particle association acts to pull the fluid together and the hard-wall contact value -- and, thus, the pressure, which is equal to $P=\lim_{z\to \sigma/2}[k_\mathrm{B}T\rho(z)]$ in the current case -- is reduced. For systems in which $\epsilon_{e_0 \cdot \mathrm{H}}=7\,k_\mathrm{B}T $, the cDFT models predict the peaks in the density profile subsequent to the contact value to be in closer proximity to the wall than the MC simulation results. The Yu-Wu model tends to underpredict while the iSAFT model tends to overpredict the magnitude of the density fluctuations, which are consequences of the averaging assumptions made in the models' mass-action laws: The Yu-Wu model evenly smears the sites across a particle's surfaces, while the iSAFT model permits sites to aggregate close to particle contacts.

\subsubsection{Modeling water near solid surfaces}

The SI demonstrates the model's ability to reproduce liquid-vapor (LV) thermodynamic data and surface tension measurements of water for temperatures up to the critical point. Model variants are compared that either incorporate the MF or HT assumptions for the dispersion interactions. Water diameters and interaction parameters were adjusted manually, starting from values that were shown to produce good fits in similar thermodynamic studies of bulk water, and are presented in Table~\ref{tab:water_parameters}. Both variants achieve excellent matches to experimental measurements of the LV densities over a broad range of temperatures, though the coexistence curve's approach of the critical point is significantly improved by the HT expansion. 

\begin{table}[t]
\caption{Parameters used to model water molecules and their interaction with soft walls. $\rho_\mathrm{w,b}$ and $\sigma_\mathrm{w}$ correspond to values at $T=300\, \mathrm{K}$. The hard-core width/diameter for graphene basal planes and the solid particles on the textured mica surface are chosen as 3.4 \AA and 1.9 \AA, respectively.}
\label{tab:water_parameters}
\begin{tabular}{l c c c c}
\hline
& MF-Yu-Wu & MF-iSAFT & HT-Yu-Wu & HT-iSAFT\\
\hline
$\rho_\mathrm{w,b}\sigma_\mathrm{w}^3$ & 0.921 & 0.921 & 0.921 & 0.921 \\
$\sigma_\mathrm{w}$ & 3.00 \AA & 3.00 \AA & 3.01 \AA & 3.01 \AA\\
$\delta$& 0.70 & 0.70 & 0.35 & 0.35\\
$\theta_\mathrm{c}$ & $26.5^\circ$ & $26.5^\circ$& $26.5^\circ$ & $26.5^\circ$\\
$(\lambda_1,\lambda_2)$ & (12,6) & (12,6) & (13,6) & (13,6)\\
$\epsilon_\mathrm{e_0\cdot H}/k_\mathrm{B}$ & 1630 K & 1630 K & 1730 K & 1730 K \\
$\epsilon^\mathrm{Mie}_\mathrm{w\cdot w}/k_\mathrm{B}$ & 270 K & 270 K & 220 K & 220 K\\
Graphene: $\epsilon^\mathrm{Mie}_\mathrm{w\cdot s}/k_\mathrm{B}$ & 41 K & 22 K & 40 K & 34 K\\
Graphene: $d_\mathrm{w\cdot s}$ & 3.43 \r{A} & 3.43 \r{A} & 3.44 \r{A} & 3.44 \r{A}\\
Mica: $\epsilon^\mathrm{Mie}_\mathrm{w\cdot s}/k_\mathrm{B}$ & 370 K & - & 520 K & -\\
Mica: $d_\mathrm{w\cdot s}$ & 2.00 \r{A} & - & 2.00 \r{A} & -\\
\hline
\end{tabular}
\end{table}

\begin{figure}[t]
\includegraphics[width=0.49\textwidth]{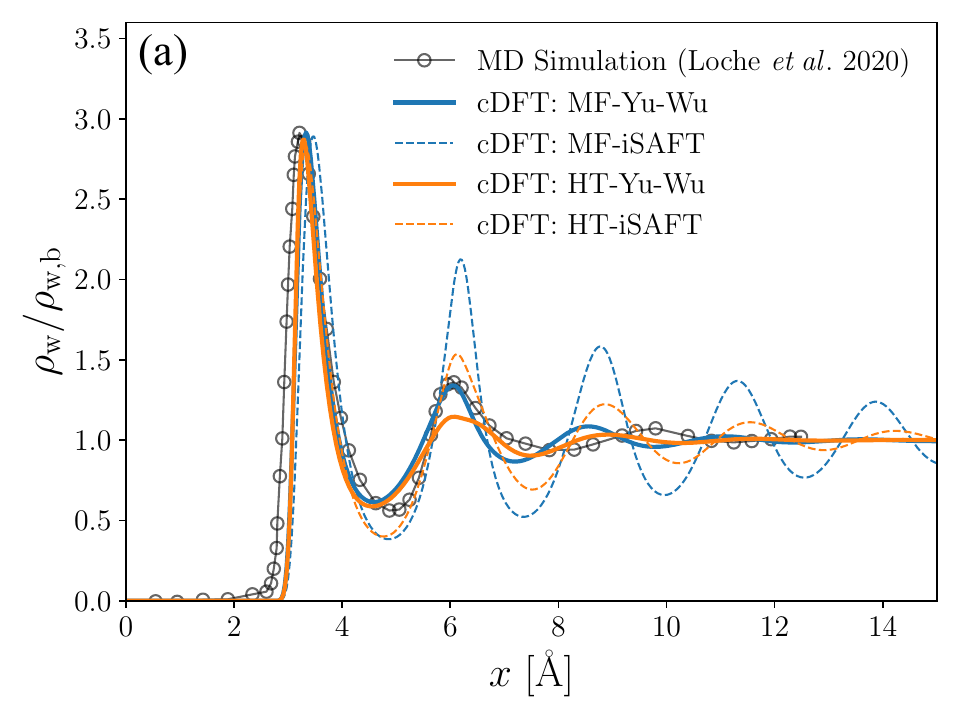}
\includegraphics[width=0.49\textwidth]{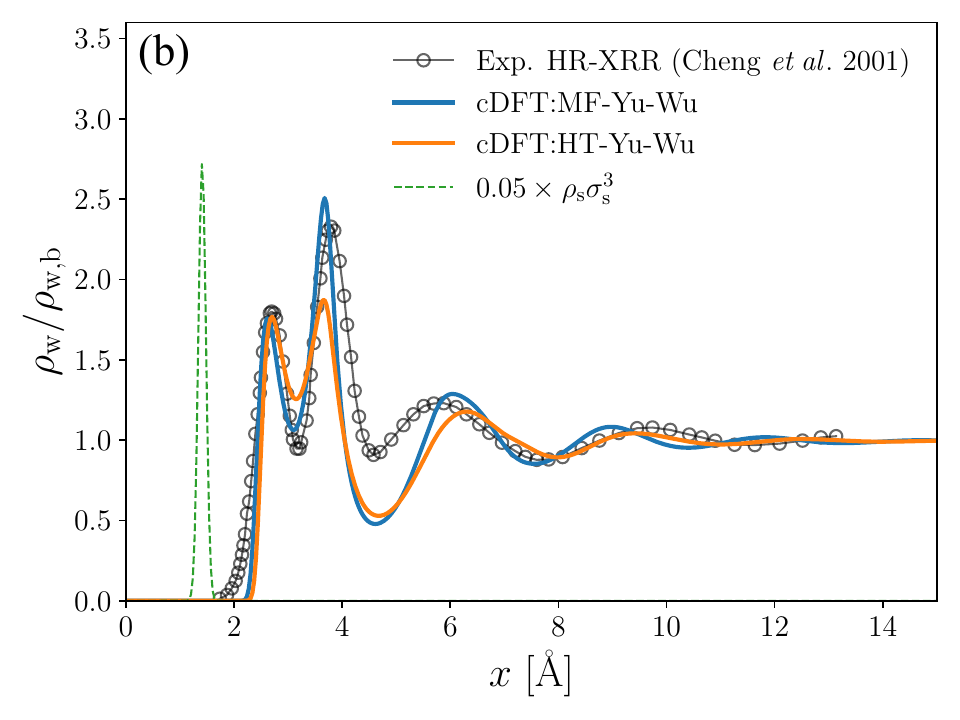}
\caption{(a) Density distribution of water molecules near a soft-wall comparing results from the MD simulations by Loche \latin{et al.}~\cite{loche2020universal} and the cDFT theory presented here; MD simulation results correspond to SPC/E water molecules confined by fixed graphene boundaries. In cDFT, the water molecules interact with the soft wall through a Steele potential, whose parameters are listed in Table~\ref{tab:water_parameters}. (b) Density distributions of water molecules near a mica surface. cDFT theory is compared to the experimentally collected high-resolution x-ray reflectivity (HR-XRR) data presented by Cheng \latin{et al.}~\cite{cheng2001molecular}. The solid boundary is simulated by explicitly introducing a fixed solid species whose scaled Gaussian density peaks are indicated by the dotted green curve.}
\label{fig:water_profiles}
\end{figure}

Next, we compare the performance of the cDFT models in resolving the density distributions of water against soft and textured surfaces. We first compare the cDFT model to MD simulations of SPC/E water against a graphene boundary~\cite{loche2020universal}. To simulate the effect of a soft boundary we use Steele's 10-4-3 potential (see Ref.(\citenum{steele1973physical}) for details),
\begin{equation}
\label{eq:Steele_potential}
V_\mathrm{ext,i}(x)=2 \pi \Delta \rho_\mathrm{s}^{V}  d_{\mathrm{s}i}^2 \epsilon_{\mathrm{s}\cdot i}^\mathrm{Mie} \left[ \frac{2}{5} \left( \frac{d_{\mathrm{s}i}}{x}\right)^{10}-\left(\frac{d_{\mathrm{s}i}}{x}\right)^4 - \frac{d_{\mathrm{s}i}^4}{3 \Delta (x+\alpha \Delta )^3}\right],
\end{equation}
which was derived by integrating a Lennard-Jones (LJ) potential for the first basal plane -- \latin{i.e.}, the first graphene plane -- and considering deeper parts of the solid to be a continuum. Above, $\rho_\mathrm{s}^{A}=\Delta\rho_\mathrm{s}^{V}$ is the areal density of the surface atoms and $\Delta$ is the basal spacing of the solid; thus, $\rho_\mathrm{s}^V$ is the volume density of the solid. Lastly, $d_{\mathrm{s}i}$ and $\epsilon_{\mathrm{s}i}$ are the LJ length and energy scales for species $i$ with the solid atoms and $\alpha=0.61$ is an empirical adjustment factor. The basal spacing for the graphene layers was chosen as $\Delta=3.4$ \r{A} and the areal density was calculated from the lattice spacing between carbon atoms as $\rho_\mathrm{s}^A = 2\sigma_\mathrm{w}^2/[\sqrt{3}(1.42 \text{ \r{A}})^2]\approx 5.15$. Values for the LJ parameters for the interaction between water and wall are listed in Table~\ref{tab:water_parameters}. The value of $\epsilon_\mathrm{s \cdot w}^\mathrm{Mie}$ was chosen to match the first peak of the density profiles for the cDFT theory and MD simulations.

The density profiles for the structure of water adjacent to the graphene boundary is plotted in Figure~\ref{fig:water_profiles}(a) for four combinations of association interaction (Yu-Wu and iSAFT) and dispersion interaction (MF and HT) schemes. It is immediately apparent that the iSAFT association scheme significantly over estimates the fluctuations in the density profile, and we choose not to further evaluate its performance in representing the structure of electrolytes below. The Yu-Wu model shows close agreement with the MD results for the magnitudes of the second and third peaks in $\rho_\mathrm{w}$. While the MF dispersion model performs better in matching amplitude of the peaks, the HT scheme better approximates the locations of the peaks. In general, the Yu-Wu cDFT models demonstrate reasonable agreement with MD simulations -- particularly considering the substantial simplifications in the molecular geometry and orientation-specific interactions -- and provide an excellent template to further explore the effect of an explicit water profile on the structure of EDLs.

Next, we look at the ability of our model to represent water density profiles near a textured surface, such as mica. The outer surface of mica is composed of silica tetrahedra, which contain cavities that are large enough for small atoms to reside in. In fact, experimental high-resolution x-ray reflectivity (HR-XRR) measurements~\cite{cheng2001molecular} and Monte Carlo simulations~\cite{park2002structure} have shown the density profile of water near mica to contain two peaks spaced $\simeq1.3$ \r{A} apart. The first peak represents adsorbed molecules that fit into the interstices of the Si tetrahedra, and the second peak is a hydration layer, in which the H$_2$O molecules hydrogen bond to surface exposed O atoms~\cite{cheng2001molecular}. 

\begin{figure}[t]
\includegraphics[width=1.0  \textwidth]{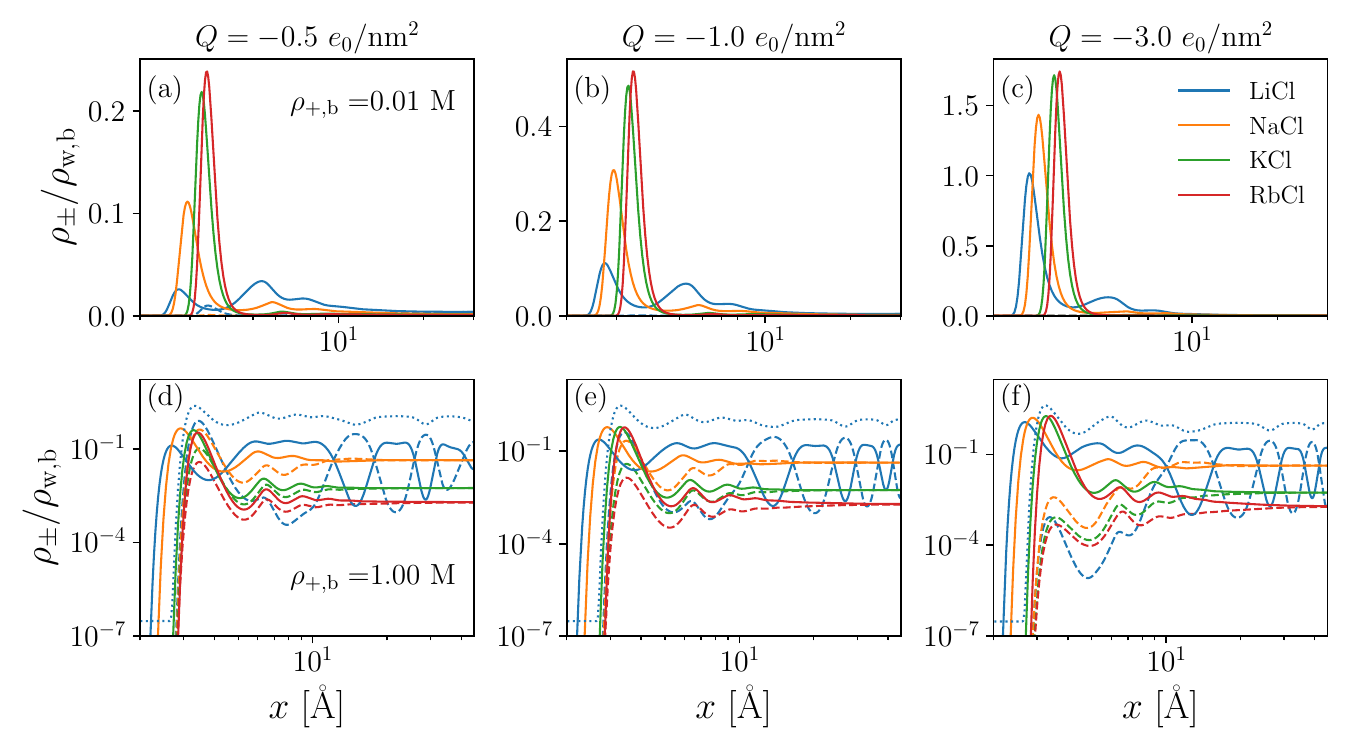}
\caption{Ion distributions near the hydrophobic interface corresponding to Figure~\ref{fig:water_profiles}(a). The top row of the figure panel (a-c) models the ion profiles for differing monovalent salts at increasing surface charge densities of (a) -0.5 $e_0/\mathrm{nm}^2$, (b) -1.0 $e_0/\mathrm{nm}^2$, and (c) -3.0 $e_0/\mathrm{nm}^2$ at a low salt concentration of 0.01 M; the $x$-axis is plotted on a logarithmic scale. The bottom row of the figure panel (d-f) corresponds to equivalent systems at a high salt concentration value of 1.00 M, where log-log axes were chosen to resolve the co-ion distribution. Solid lines indicate counter-ion profiles, dashed lines indicate co-ion profiles, dotted lines indicate water molecule profiles for the LiCl system. Table~\ref{tab:ion_parameters} lists the hard-sphere diameter and dispersion energy parameters chosen for each salt.}
\label{fig:monovalent_hydrophobic}
\end{figure}

\begin{figure}[t]
\includegraphics[width=1.0  \textwidth]{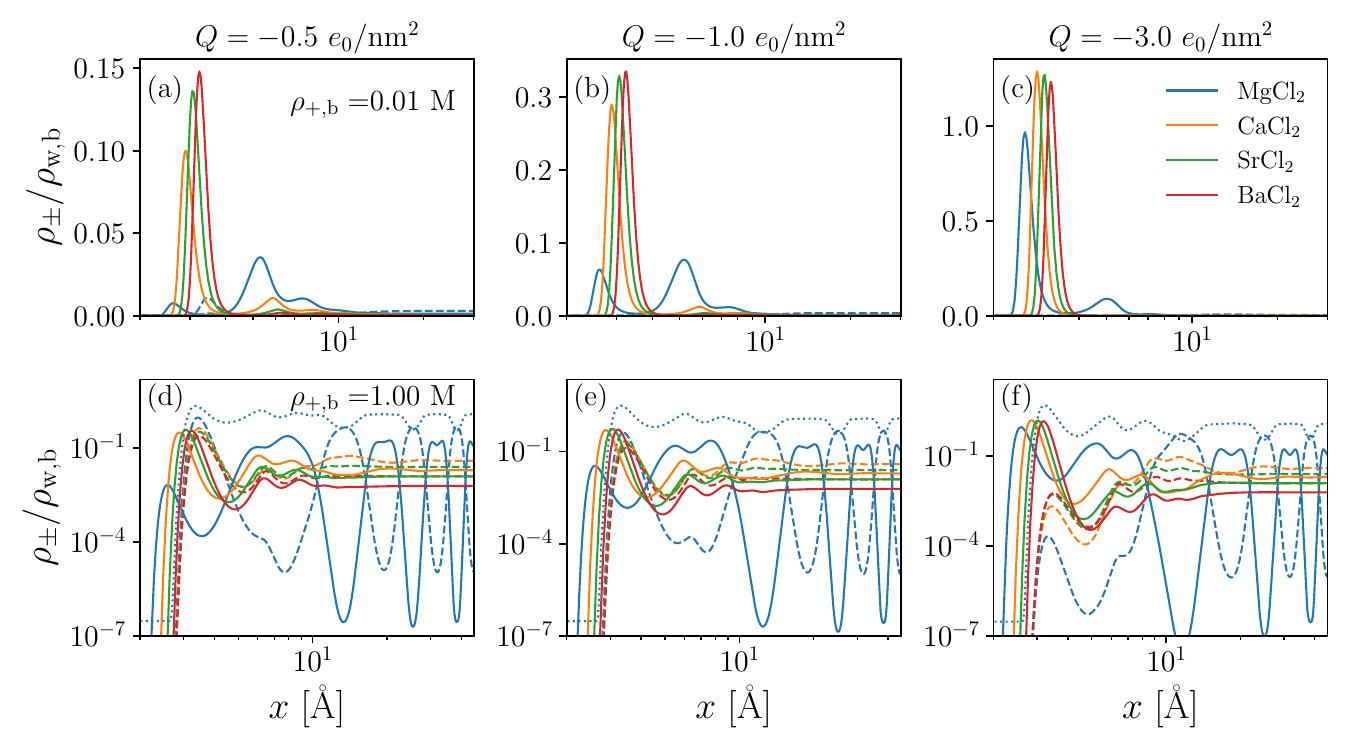}
\caption{Ion distributions for divalent salts near a hydrophobic interface. Panels are organized as in Figure~\ref{fig:monovalent_hydrophobic} and Table~\ref{tab:ion_parameters} lists the diameters and interaction energies for the ions.}
\label{fig:divalent_hydrophobic}
\end{figure}

To take the surface texture of mica into account, we represent the outer most solid layer explicitly by prescribing its density profiles within the simulation domain. We place the solid layer at 1.28 \r{A} to match the separation between peaks measured experimentally. The layer is represented by a layer of oxygen atoms of hard-sphere diameter $\sigma_\mathrm{O}=1.2$ \r{A} that have a narrow Gaussian distribution with areal density $\rho_\mathrm{O}=6 \sigma_\mathrm{w}^2 / (5.19 \text{ \r{A}} \times 9.01 \text{ \r{A}}) \approx 1.16$ -- the density of surface oxygen atoms on an exposed tetrahedral sheet~\cite{jackson1931crystal}. The solid oxygen layer is indicated by the green dotted curve in Figure~\ref{fig:water_profiles}(b). Explicit representation of the surface atoms endows them with FMT densities and permits them to hydrogen bond with water molecules and cations. Thus, we re-write the mass action laws in Eqn.(\ref{eq:mass_action_YuWu}b,c) as
\begin{subequations}
\begin{align}
& \chi_\mathrm{H}^\mathrm{w} + 2n_{0,\mathrm{w}}\zeta_\mathrm{w} \chi_{e_0}^\mathrm{w}\chi_{\mathrm{H}}^\mathrm{w} \Delta_{e_0\cdot \mathrm{H}} + M_\mathrm{-} n_{0,\mathrm{-}}\zeta_\mathrm{-} \chi_\mathrm{H}^\mathrm{w}\chi^{\mathrm{-}} \Delta_{\mathrm{H}\cdot \mathrm{-}} + n_{0,\mathrm{O}}\zeta_{\mathrm{O}}\chi_\mathrm{H}^{\mathrm{w}}\chi_{e_0}^{\mathrm{O}} \Delta_{e_0\cdot \mathrm{H}}= 1,\\
& \chi^{+} + M_+ n_{0,\mathrm{w}}\zeta_\mathrm{w} \chi_{e_0}^\mathrm{w}\chi^\mathrm{+} \Delta_{e_0\cdot \mathrm{+}}+ n_{0,\mathrm{O}}\zeta_{\mathrm{O}}\chi^{+} \chi_{e_0}^{\mathrm{O}} \Delta_{e_0 \cdot +} = 1.
\end{align}
\end{subequations}
and the fraction of hydrogen bonded surface atoms is calculated from
\begin{equation}
\chi_{e_0}^\mathrm{O} + n_{0,\mathrm{w}}\zeta_{\mathrm{w}}\chi_{e_0}^{\mathrm{O}}\chi_\mathrm{H}^{\mathrm{w}} \Delta_{e_0 \cdot \mathrm{H}} + n_{0,+}\zeta_{\mathrm{w}}\chi_{e_0}^{\mathrm{O}}\chi^+ \Delta_{e_0 \cdot +} =1.
\end{equation}
Here, each surface oxygen atom has one bonding site and an association strength we naively set equal to that of the oxygen atoms on water molecules bonding to the hydrogen atoms on neighboring water molecules, $\Delta_{e_0 \cdot \mathrm{H}}$, or neighboring cations, $\Delta_{e_0 \cdot +}$. Similarly, the interaction strength for the van der Waals interactions is set to that between water molecules, $\epsilon_\mathrm{w\cdot O}=\epsilon_\mathrm{w\cdot w}$. To account for the solid structure in mica beneath the surface atoms, a Steele potential (Eq.(\ref{eq:Steele_potential})) is added as was done for the graphene interface.

The parameters used to model the water-Mica interface are listed in Table~\ref{tab:water_parameters}.  As before, we set $\epsilon^\mathrm{Mie}_\mathrm{s\cdot \mathrm{w}}$ in $V_\mathrm{ext}$ to the value that matches the first peak in the water profiles. Figure~\ref{fig:water_profiles}(b) shows that the explicit solid layer assists in recreating the two surface peaks -- absorption and first hydration layers. While the MF theory does better to match the amplitudes of the peaks, the HT theory more accurately identifies the locations of the peaks further removed from the interface. Both theories perform poorly in predicting the density profile near the trough following the first hydration layer; this is certainly due, in significant part, to our simplified model for the surface interactions. Though the statistics near the planar surface depend only on $x$, proper modeling of the texture requires two- or three-dimensional cDFT~\cite{martin2016atomically}. Nonetheless, the planar profiles should be adequate in informing how the structure and interactions of water near a mineral interface contribute to the distribution of ions within an EDL.

\subsubsection{Modeling ion distributions near charged surfaces}

\begin{table}
\caption{Ion parameters taken from Eriksen \latin{et al.}~\cite{eriksen2016development} and, where necessary, calculated from combining rules. Dispersion interactions between cation and anions are relatively small -- compared to the Coulomb interaction energy -- and are neglected in this study.}
\label{tab:ion_parameters}
\begin{tabular}{l c c c c}
\hline
& $\sigma_i$ [\r{A}] & $\epsilon_{\mathrm{w} \cdot \pm}^\mathrm{Mie}/k_\mathrm{B}$ [K] & $\epsilon_{\mathrm{s} \cdot \pm}^\mathrm{Mie}/k_\mathrm{B}$ [K] & $M_{\pm}$\\
\hline
Li$^+$ & 1.80 & 1023 & 1402 & 5 \\
Na$^+$ & 2.23 & 540 & 739 & 6 \\
K$^+$ & 3.04 & 376 & 515 & 8 \\
Rb$^+$ & 3.32 & 354 & 485 & 10 \\
Mg$^{2+}$ & 1.72 & 2235 & 3063 & 6 \\
Ca$^{2+}$ & 2.28 & 1461 & 2000 & 6 \\
Sr$^{2+}$ & 2.64 & 1047 & 1435 & 8 \\
Ba$^{2+}$ & 2.98 & 840 & 1151 & 8 \\
Cl$^-$ & 3.34 & 95 & - & 6 \\
OH$^-$ & 2.46 & 134 & 184 & 3\\
\hline
\end{tabular}
\end{table}

Predictions for the structures of EDLs with monovalent electrolytes are shown in Figure~\ref{fig:monovalent_hydrophobic}. In all systems, we choose the water-graphene interface that was modeled using the Yu-Wu association scheme and MF dispersion interactions as a template. The solid is subsequently charged to surface charge densities of $Q=-0.5\,e_0/\mathrm{nm}^2$, $Q=-1.0\,e_0/\mathrm{nm}^2$, or $Q=-3.0\,e_0/\mathrm{nm}^2$. The indicated alkali ions differ in their size and inter-water dispersion energies; these values are taken from Eriksen \latin{et al.}~\cite{eriksen2016development} and are listed in Table~\ref{tab:ion_parameters}. Additionally, we set the dispersion energy between the ions and the solid boundary to the same value chosen for the interaction between the water molecules and the solid boundary, $\epsilon^\mathrm{Mie}_{\mathrm{w\cdot s}}=\epsilon^\mathrm{Mie}_{\mathrm{+\cdot s}}=\epsilon^\mathrm{Mie}_{\mathrm{-\cdot s}}$, and ensure that the bulk packing density of the electrolyte matches the bulk packing density of the salt-free system. At low salt concentration (0.01 M; Figure~\ref{fig:monovalent_hydrophobic}(a-c)), Na$^+$, K$^+$, and Rb$^+$ form a single layer that compensates for graphene's surface charge. The small Li$^+$ ions, on the other hand, form multiple layers, even showing a small co-ion layer for $Q=-0.5\,e_0/\mathrm{nm}^2$. As the surface charge increases, the Cl$^-$ co-ions are pushed out of the domain and the distribution of the Li$^+$ ions shifts closer to the first peak near the surface. The difference in the trends for LiCl relative to the other monovalent salts is largely due to the strong interaction energy between Li$^+$ and water.

At high salt concentration and low surface charge density (1.00 M and $Q=-0.5\, e_0/\mathrm{nm}^2$; Figure~\ref{fig:monovalent_hydrophobic}(d)), all salts produce strong co-ion peaks that reside just outside the space occupied by the first counter-ions layer. These peaks are due to the hydrophobic nature of the interface that pushes the Cl$^-$ ions to the interface. The Li$^+$ system predicts alternating counter-ion and co-ion peaks with the largest counter-ion peak located 0.8 \r{A} from the graphene surface. It appears that the structuring of the lithium density profiles is due to its strong hydration forces, which -- at high salt concentration -- compete with steric, excluded volume interactions to produce preferential regions of high concentration. Also shown in Figure~\ref{fig:monovalent_hydrophobic}(d-f) are the water density profiles associated with the LiCl systems, which show strong correlation with the counter-ion distribution. For all salts, the co-ion peaks near the solid boundary are increasingly suppressed as the surface charge density is increased.

\begin{figure}[t]
\includegraphics[width=0.49  \textwidth]{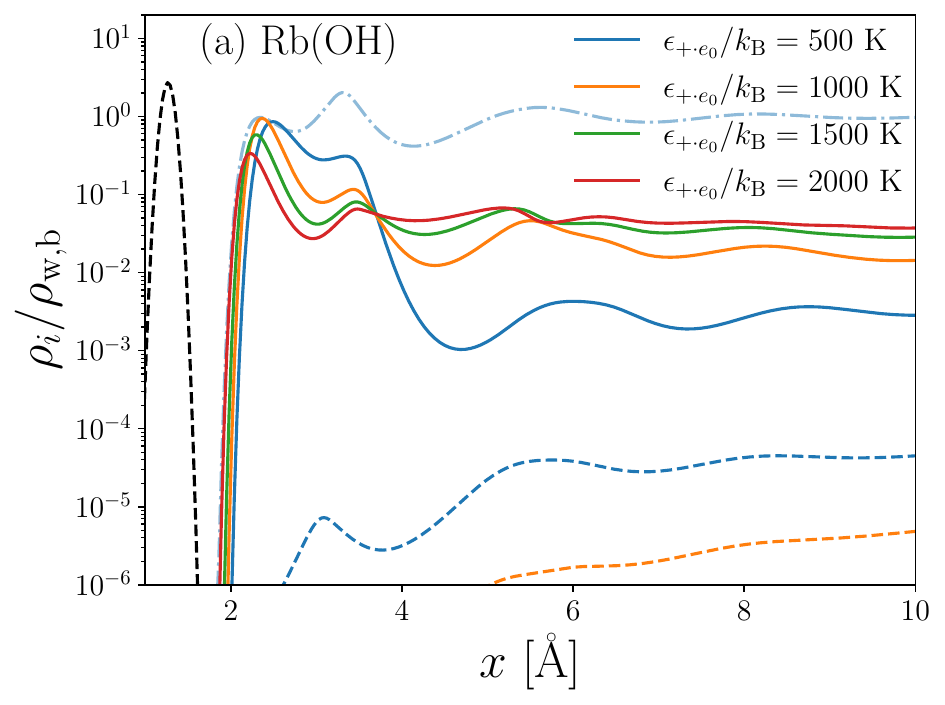}
\includegraphics[width=0.49  \textwidth]{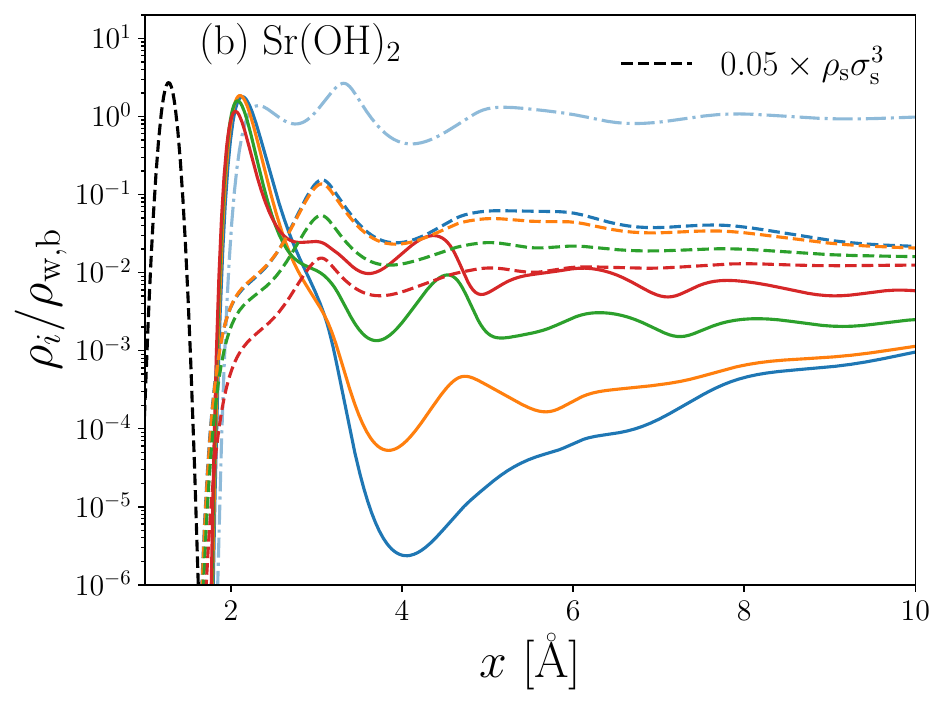}
\caption{Ion distributions near the hydrophilic, textured mica interface corresponding to Figure~\ref{fig:water_profiles}(b). The left panel (a) shows the distribution of monovalent rubinium counter-ions and the right panel (b) shows the distribution of divalent strontium counter-ions. Colors correspond to different associative interaction energies, which are applied equally to cation-water and anion-water association, $\epsilon_{+\cdot e_0}=\epsilon_{-\cdot \mathrm{H}}$. Also shown in the panels are the distributions of co-ion (dashed lines), water molecules (opaque blue dash-dotted lines; only for $\epsilon_{+\cdot e_0}=500\, \mathrm{K}$), and the first layer of the solid surface.}
\label{fig:mica_ion_distribution}
\end{figure}

For completeness, we also show the predicted density profiles for divalent salt systems in Figure~\ref{fig:divalent_hydrophobic}. Interestingly, despite having dispersion interaction energies that exceed those of Li$^{+}$, the divalent cation systems containing Ca$^{2+}$ and Sr$^{2+}$ do not show the alternating layering between counter- and co-ions. This suggests that the small size of the Li$^{+}$ and Mg$^{2+}$ ions plays a role in their anomalous structuring near the charged interface: Approximately two of either cation fits in the space of one water molecule. This permits the dispersion interactions, modeled using a radially symmetric and orientation-independent interaction energies, to minimize $\mathcal{F}_\mathrm{att}$ by organizing into co-located regions of high $\rho_+$ and $\rho_\mathrm{w}$. Similar layering due to the competition between steric repulsion and electrostatic ineteractions has been modeled for room-temperature ionic liquids~\cite{kirchner2013electrical,de2020interfacial}. Certainly, the solvation of ions strongly orientates the neighboring water molecules, which is currently unaccounted for. We explore the effect of adding associative interactions between ions and water molecules along a mica-electrolyte interface next -- although not explicitly accounting for orientation, the statistics of the molecules' bonding sites places limits on the number and types of nearest-neighbors.

Figure~\ref{fig:mica_ion_distribution} shows the distribution of (a) Rb(OH) or (b) Sr(OH)$_2$ bases near the mica interface. The dispersion interaction energy between the implicit solid boundary (modeled using Steele potentials) and the ions is estimated using the Berthelot combining rule: $\epsilon_\mathrm{s\cdot\pm} = \sqrt{(\epsilon_\mathrm{w\cdot s}^{\mathrm{Mie}})^2 (\epsilon_\mathrm{w\cdot \pm}^{\mathrm{Mie}})^2 / (\epsilon_\mathrm{w\cdot w}^{\mathrm{Mie}})^2}$; their values are recorded in Table~\ref{tab:ion_parameters}. The surface charge is set to $-1\,e_0/A_\mathrm{uc}$, where $A_\mathrm{uc}=46.7$ \r{A}$^2$ is the area per unit cell, and the bulk concentration of the cations is $\rho_\mathrm{+,b}=0.01$ M. We treat the explicit solid layer, which represents the exposed oxygen layer of the silica tetrahedra, by setting the dispersion energy to those between the ions and water molecules and permit associative bonding with cations. Until now, association between water molecules and ions was neglected. Figure~\ref{fig:mica_ion_distribution} shows the effect of increasing the association strength, measured by $\epsilon_{+ \cdot e_0}$, between water molecules and cations (the same value is chosen for the association between anions and water molecules and cations and the surface oxygen atoms). A study investigating the ability of association and dispersion interactions to represent the vapor pressure, densities, and activity coefficients of bulk NaCl solutions was conducted by Gil-Villegas \latin{et al.}~\cite{gil2001statistical}. Although their high-temperature expansion of the dispersion interactions produced better fits over a range of salt concentrations, under our current formulation, association interactions permit bonds to be formed between a restricted set of nearest-neighbors; this is particularly relevant when trying to capture surface adsorption at a finite number of sites. Thus, a combination of both association and dispersion interactions may prove to be most effective.

\begin{figure}[t]
\includegraphics[width=1.0  \textwidth]{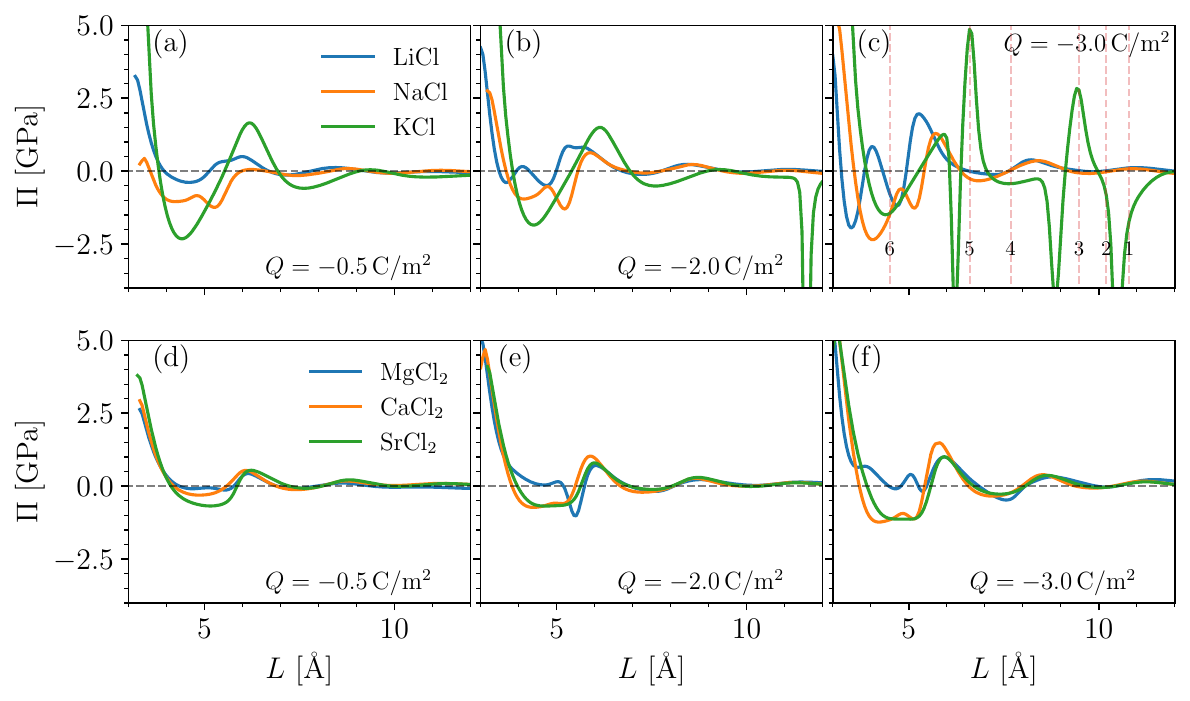}
\caption{cDFT predicted disjoining pressure between charged graphene surfaces plotted for electrolytes composed of (a-c) monovalent, akali metal chloride solutions and (d-f) divalent, alkaline earth chloride solutions as a function of surface separation, $L$. The surface charge increases from -0.5 C/m$^2$ to -2.0 C/m$^2$ to -3.0 C/m$^2$ moving from left to right, and all systems have a salt concentration of 1.0 M. The vertical, dashed red lines in panel (c) correspond to the surfaces separations plotted in Figure~\ref{fig:pressure_potassium_closer_look}.}
\label{fig:graphene_disjoining_pressure}
\end{figure}

At low $\epsilon_{+ \cdot e_0}$, ions aggregate close to the interface. As $\epsilon_{+ \cdot e_0}$ is increased, the solvation shell around the ions brings water molecules closer to the interface and spreads the counter-ions into a more diffuse layer. The dual peaks of for the RbOH system show similarity to MD simulation results modeling the Stern layer of akali ions near mica surfaces~\cite{bourg2017stern}. MD simulations show and experiments suggest that the first peak represents ions sitting in the ditrigonal cavities between silcia tetrahedral and the second peak corresponds to ions sitting above the ionized surface atoms~\cite{bourg2017stern,park2006hydration}. Certainly the details of the $yz$-structuring of the cations cannot be fully accounted for in the 1D cDFT theory presented here. Nonetheless, explicitly incorporating the oxygen atom density peaks permits some control over the competition between adsorption of water molecules and ions to the charged surface sites. Better results are yet expected if the orientations of the water molecules can be included and correlated to their ability to hydrogen bond; as currently constructed, bonding sites are smeared over the surface area of water and ion species, which allows, for instance, the same water molecule to make hydrogen bonds with the solid surface on the left and anions on the right.

For the Sr(OH)$_2$ system, charge reversal is observed with the formation of a co-ion layer that peaks at a distance of $x=3.0$ \r{A} from the boundary. Similar charge reversal was observed in MD simulations of muscovite mica in contact with a concentrated NaCl solution of 0.5 M~\cite{sakuma2011structure}.  Upon increasing the association energy between ions and water molecules, the peak of the cation monolayer adjacent to the surface decreases and the second, co-ion layer peak also decreases. Nonetheless, divalent ions tend to overscreen the surface charge, which, as we shall see, has implications for the disjoining pressure.

\begin{figure}[t]
\includegraphics[width=1.0\textwidth]{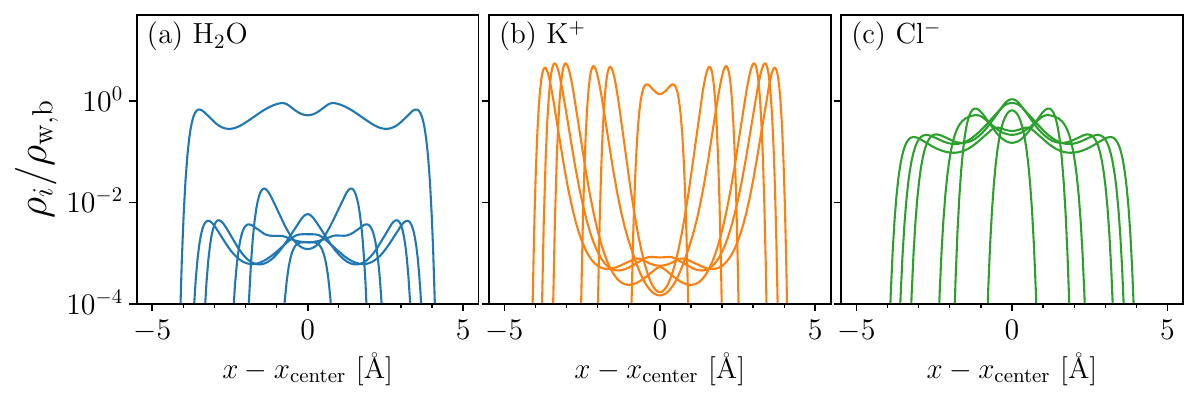}
\caption{Water (a; blue), cation (b; orange), and anion (c; green) density profiles plotted between two impinging graphene surfaces. The profiles correspond to a charge density of $Q=-3.0$ C/m$^2$ and the surface separations are indicated by the red, vertical lines in Figure~\ref{fig:graphene_disjoining_pressure}(c).}
\label{fig:pressure_potassium_closer_look}
\end{figure}

\subsubsection{Disjoining pressure between charged surfaces}
As a final task, the inter-surface pressure between impinging graphene or impinging mica surfaces is calculated. This thermodynamic stress -- also known as the disjoining pressure -- is readily accessed by measuring the free energy changes in an open system (Eq.(\ref{eq:grand_potential})) that accompany changes in the system's volume,
\begin{equation}
\Pi = - \left(\frac{\partial \Omega}{\partial V}\right)_{\mu,T} - P_\mathrm{b},
\end{equation}
where $P_\mathrm{b}=-f_\mathrm{b}(\{\rho_\mathrm{i,b}\}) + \sum_{i} \rho_{i,b} \mu_i$ is the pressure in the bulk, with $f_\mathrm{b}$ the homogeneous bulk free energy density. 

In calculating $\Pi$ we modify the fMSA treatment for the electrostatic correlations in Eq.(\ref{eq:electrostatic_energy_density}). We set the capacitance radius of the ions to their hard-sphere radius, $b_i=\sigma_i/2$. This was found to be necessary to avoid interference of the shells at one interface with the opposing EDL, and is justified by the fact that the $b_i$ should shrink as the local ion concentrations and those in their vicinity are less and less represented by the bulk values. A comparison of the performance of the cDFT model to reproduce MC simulation results of poly-disperse counter- and coion distributions near a charged hard-wall are shown in Figure (S4). The system corresponds to a dilute solution of ions in a dielectric continuum neglecting any explicit solvent interactions, and, thus, has comparatively broader density peaks. Nonetheless, it is readily apparent that inclusion of the fMSA electrostatic free energy significantly improves the correspondence of MC and cDFT results, regardless of the choice for $b_i$. While use of the bulk capacitance radius provides a better comparison in dilute conditions near a single charged boundary, the two solutions should converge as the concentration is raised.

As displayed in Figure~\ref{fig:graphene_disjoining_pressure}, approaching graphene surfaces produce larger oscillations in $\Pi$ for monovalent salts than they do for divalent salts. In fact, the pressures for the divalent system (Figure~\ref{fig:graphene_disjoining_pressure}(d-f)) seem to be fairly insensitive to the surface charge density, showing modest fluctuations that form shallow minima around $x=4.5$-$5.5$ \r{A}. In general, the divalent systems maintain a stricter organization of the ions -- that is, ions that are more tightly bound -- and pressure oscillations relate directly to the continuous elimination of water layers. The small size of the magnesium ions disrupts these oscillations. Due to its strong dispersion interactions with water, the MgCl$_2$ solution produces two peaks (as seen in Figure~\ref{fig:divalent_hydrophobic}) that coalesce as the graphene surfaces are brought together.

The monovalent systems, on the other hand, form pressure oscillations that are both larger in magnitude and more irregular. Particularly intriguing are the attractive and repulsive spikes in $\Pi$ with the KCl-containing electrolyte at high surface charge density, $Q=-3.0\,\mathrm{C/m^2}$. We inspect the water, cation, and anion density profiles for the KCl solution more closely in Figure~\ref{fig:pressure_potassium_closer_look}(a-c), where we plot the distributions for the surface separations marked and numbered in Figure~\ref{fig:graphene_disjoining_pressure}(c). Intriguingly, the divergences in the pressure relate to the dehydration of the interlayer space and subsequent crowding of co-ions. As the surfaces approach, the interlayer space is intially filled with hydrated cations (1). As the surfaces move together (1$\to$2), the cations dehydrate, forcing the water molecules into the bulk and replacing them with the Cl$^-$ anions. At this point, the gap between the graphene layers is essentially composed of an ionic liquid. This leads to a strong inter-surface attraction, until the Cl$^-$ ions in the center become crowded (3) and are forced out of the space in resistance of their electrostatic forces with the K$^+$ ions (3$\to$4). Eventually, the anions collapse into a single layer between cationic Stern layers (5$\to$6). These interesting dynamics are due, in part, to K$^+$'s relatively low affinity for water. As the graphene surfaces are not particularly water loving, strong solvation forces are required by the ions to keep water molecules from retreating into the bulk. This is satisfied for the LiCl and NaCl systems: Here, the more modest fluctuations in pressure result from the competition between steric repulsion and the inter-species association and dispersion interactions that aim to maintain the the water coordination.

Diao \latin{et al.} measured the force exerted between two graphene layers saturated in either NaCl or KCl solution at different molarity using an atomic force microscope (AFM)~\cite{diao2019slippery}. Albeit being measured at charges significantly lower than those in Figure~\ref{fig:graphene_disjoining_pressure}, their experiments showed distinguishable steps in the force curve -- representative of troughs in the $\Pi$ vs. $L$ curve. Measurements made in the KCl solution displayed a single step near the surface, corresponding to the displacement of the surface solvating water layer, and multiple shorter steps for measurements made in the NaCl solution. Similar observations hold true for our cDFT predictions, where the increased oscillations trace the multiple hydration states of the Li$^{+}$ and Na$^{+}$ that are elicited as the surfaces are brought together.

\begin{figure}[t]
\includegraphics[width=1.0\textwidth]{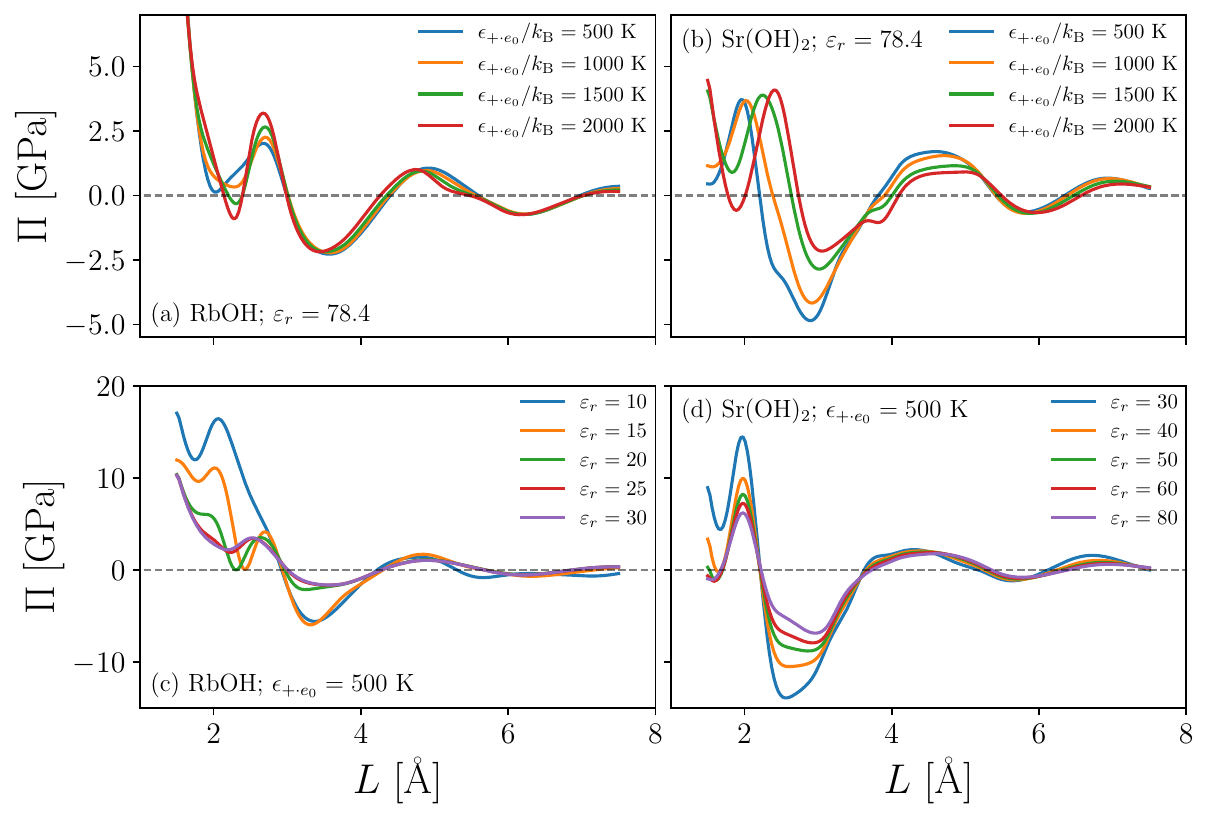}
\caption{cDFT predicted disjoining pressure between charged mica surfaces plotted for electrolytes containing (a) RbOH and (b) Sr(OH)$_2$ as a function of surface separation, $L$. The differently colored lines correspond to different prescribed association strengths between cations and water molecules/surface oxygen sites. The surface charge is set to $Q=-1.0 e_0/A_\mathrm{uc}=-2.14 e_0/\mathrm{nm}^2$ with $A_\mathrm{uc}=46.7$ \r{A}$^2$ and $\mathrm{\rho}_\mathrm{+,b}=0.01\,\mathrm{M}$. The bottom panels plot the disjoining pressure for systems containing (c) RbOH and (d) SR(OH)$_2$ electrolytes for different values solvent permittivity, $\varepsilon_\mathrm{r}$, while holding $\epsilon_{+\cdot e_0}/k_\mathrm{B}=\epsilon_{-\cdot H}/k_\mathrm{B}$ fixed at 500K.}
\label{fig:mica_pressure}
\end{figure}

As a last effort, we investigate the disjoining pressure between hydrophilic mica surfaces separated by solutions containing Rb$^+$ and Sr$^{2+}$ cations (see Figure~\ref{fig:mica_pressure}). As noted earlier, the Sr$^{2+}$ system overscreens the surface charge and produces an outer boundary region prevalent in co-ions. When this region is separated by two Stern layers composed of divalent ions a highly inter-attractive stress is produced that reaches as high as 5 GPa at surface separation of about 3 \r{A} (as measured between the centers of the surface oxygen sites; Figure~\ref{fig:mica_pressure}b)). Similarly strong pressures are reported between C-S-H surfaces modeled using MD~\cite{abdolhosseini2014combinatorial,goyal2021physics}. Interestingly, Goyal \latin{et al.} identified the structuring of water molecules within the cationic Stern layer and the resulting ion correlations as a key mechanism for the C-S-H cohesion~\cite{goyal2021physics}. In the authors' model, water molecules and ions organize along atomically flat, charged surfaces. What we find in the current article is that the texture of the solid and its non-electrostatic interactions with the water molecules and ions plays an important role in the inter-surface pressure. Specifically, overscreening can lead to excess electrostatic forces that cause strong $\Pi$. This is easily demonstrated by investigating the effect of added ion-water association. As the association forces of the Sr$^{2+}$ and OH$^-$ ions increase -- remembering that we set $\epsilon_{+\cdot e_0}=\epsilon_{-\cdot \mathrm{H}}$, and allow association between water and the exposed oxygen groups to grow -- the magnitude of the attractive well reduces. In fact, for $\epsilon_{+\cdot e_0}=2000\,\mathrm{K}$ the $\Pi$ vs. $L$ curve mimics that of the monovalent RbOH system. Thus, the additional hydration forces require larger pressures to displace the water molecules and help suppress the co-ion layer, which otherwise contributes to charge reversal and the buildup of additional electrostatic forces between opposing EDLs.

Under strong coupling between ions and the electrostatic field of charged surfaces, ions are known to organize into a Wigner lattice in the Stern layer, which can produce significant like-charge attraction~\cite{moreira2000strong,naji2005electrostatic}. The 1D model presented here does not capture in-plane ion correlations, such that the strong inter-surface attraction arises entirely from the layered structuring of counter- and co-ions and their association with the solvent. This is demonstrated in Figure (S5) in the Supporting Information, where we plot the disjoining pressure as a function of surface separation for dilute electrolytes in an implicit solvent. The observed inter-surface attraction is significantly lower than what would be predicted from Monte Carlo simulations of point-charges or charged hard-spheres~\cite{moreira2000strong,vsamaj2020strong}. 

It is known that the dielectric permittivity field of water is anisotropic and oscillates near interfaces~\cite{bonthuis2011dielectric,goyal2021physics,de2022polar}; if the surface charge is significant enough, the perpendicular component of dielectric response function displays singularities and is negative, indicative of overscreening~\cite{bonthuis2011dielectric}. To probe the sensitivity of our model to changes in the permittivity of the solvent, Figure \ref{fig:graphene_disjoining_pressure}(c) and \ref{fig:graphene_disjoining_pressure}(d) plot $\Pi$ as a function of $L$ for the previously described RbOH and Sr(OH)$_2$ systems while holding $\epsilon_{+\cdot e_0}=\epsilon_{-\cdot H}$ constant at 500K and varying $\varepsilon_\mathrm{r}$. For low values of the permittivity, ions in the Stern layer are bound more tightly to the charged interface, requiring larger pressures to displace the particles for $L \lesssim 2.5$. At larger separation, the bound ions, together with the interpenetrating coions, assist in holding the surfaces together, further reducing the minimum in $\Pi$.

\section{Conclusions}

In this work, we employed cDFT to model the structure of ions in an explicit solvent near charged interfaces. We started by carefully developing a model of water molecules that reasonably reproduces the LV coexistence and the surface tension between the phases. Within this process, we compared two association schemes and two schemes for the long-range dispersion interactions. In comparison to the iSAFT association scheme, the Yu-Wu model showed significant improvement in approximating the density profiles of water molecules near hard walls, while both the MF and HT models for the long-range pair correlation function showed good representation of the thermodynamic properties. Additionally, we incorporated electrostatic correlations through a fMSA that has shown previous success in resolving concentration profiles in implicit solvent models. It is acknowledged that much has been done over the years to fine-tune similar bulk thermodynamic and cDFT models to reproduce the physical properties of solvents and polymeric systems~\cite{gil1997statistical,eriksen2016development,pereira2011integrated}. As such, the aim of the current study was to assemble these developments to investigate the structuring of ions close to textured surfaces. Remarkably, inclusion of hard-sphere, association, dispersion, and electrostatic interactions provides a rich parameter space that permits modeling the molecular layering of solvent and ions near charged interfaces.

Many materials of industrial importance have surfaces whose charge depends on the deprotonation or ionization of the surface groups; water molecules (or the solvent) help regulate this charge and further template the interface, impacting the distribution of mobile ions~\cite{gonella2021water,goyal2021physics}. This organization has important implications for the disjoining pressure between, \textit{e.g.}, clay particles in soils, C-S-H in cement, and electrodes in batteries. Within this context, we demonstrated several interesting features near implicitly and explicitly resolved solid surfaces. 
\begin{itemize}
\item In cases where the electrolyte between surfaces has a high salinity and the ions and surface molecules maintain weak hydration forces, the approach of charged surfaces causes water molecules to be expelled from the region, after which subsequent ion layering produces alternating spikes in pressure. These spikes in pressure arise from the strong electrostatic forces between counter- and co-ions. This phenomena should be investigated in greater detail, as the current model does not account for the change in the dielectic constant that must accompany the absence of the water molecules. Even near isolated interfaces, it is known that the polarization of water exhibits fluctuations over approximately a nanometer~\cite{tielrooij2009dielectric,bonthuis2011dielectric}
\item In the case of divalent salts near hydrophilic surfaces, it was shown that overscreening can lead to strongly inter-attractive pressures that approach 5 GPa in magnitude. A necessary ingredient for the high inter-attraction is local charge-reversal, which acts to pull the surface-adsorbed counter-ions toward the center of the gap. The origin of this inter-attraction is thus different from that found in systems purely composed of counter-ions and solvent, where it has been shown that in-plane ion-ion correlations can lead to a strong attraction between similarly charged surfaces~\cite{goyal2021physics,pegado2008like,lee2021ion}.
\end{itemize}

While many of the measured quantities require further model development to warrant reliable prediction of salt-specific electrolyte behavior, the current study -- at a minimum -- provides a thorough approach to mapping trends in the density profiles of aqueous EDLs. Though not emphasized here, the model may readily be calibrated to experimental data on the pressure and activity coefficients of electrolyte systems to lend further credence to calculated bulk thermodynamic quantities~\cite{eriksen2016development}. More challenging is the work needed to faithfully incorporate hydration interactions at charged solid interfaces and the modifications in the dielectric constant that come from correlations in the water dipoles~\cite{conway1951dielectric}. In principle surfaces such as mica have sites that reorient water molecules to permit association. These details cannot be resolved by the Yu-Wu association model, which smears sites across the surface area of the molecules. Additionally, the present study evaluated one-dimensional structuring of fluids along solid surfaces that have a two-dimensional surface texture. Of course, important details in the texture are lost in averaging the solid density across the $yz$-plane: The same assumptions of ergodicity bestowed the fluid do not hold. Thus, some studies have extended simpler cDFT models to two-dimensional surface textures~\cite{martin2016atomically}. Our ongoing work is dedicated to improving some of these shortcomings.  

%%%%%%%%%%%%%%%%%%%%%%%%%%%%%%%
%% ACKNOWLEDGEMENT %%
%%%%%%%%%%%%%%%%%%%%%%%%%%%%%%%

\begin{acknowledgement}

This work was supported by the author's start-up funding from the University of Southern California's Viterbi School of Engineering. The author thanks 
Prof. Mohammad Javad Abdolhosseini Qomi at the University of California at Irvine for an insightful discussion during the American Society of Civil Engineering's 2023 Engineering Mechanics Institute Conference in Atlanta, GA and the manuscript reviewers, who offered insightful comments and discussion points.

\end{acknowledgement}

\begin{suppinfo}
Additional information on calculating $c_1$ and $c_2$ in Eq.(\ref{eq:effective_packing}), a description of the iSAFT association model, simplified expressions for the association and dispersion interaction components of the single-particle direct correlation function $\delta \mathcal{F}_\mathrm{ex}/\delta \rho_i$, and Figures S1-S5 with supporting text.
\end{suppinfo}

\bibliography{references}

\providecommand{\latin}[1]{#1}
\makeatletter
\providecommand{\doi}
  {\begingroup\let\do\@makeother\dospecials
  \catcode`\{=1 \catcode`\}=2 \doi@aux}
\providecommand{\doi@aux}[1]{\endgroup\texttt{#1}}
\makeatother
\providecommand*\mcitethebibliography{\thebibliography}
\csname @ifundefined\endcsname{endmcitethebibliography}  {\let\endmcitethebibliography\endthebibliography}{}
\begin{mcitethebibliography}{91}
\providecommand*\natexlab[1]{#1}
\providecommand*\mciteSetBstSublistMode[1]{}
\providecommand*\mciteSetBstMaxWidthForm[2]{}
\providecommand*\mciteBstWouldAddEndPuncttrue
  {\def\EndOfBibitem{\unskip.}}
\providecommand*\mciteBstWouldAddEndPunctfalse
  {\let\EndOfBibitem\relax}
\providecommand*\mciteSetBstMidEndSepPunct[3]{}
\providecommand*\mciteSetBstSublistLabelBeginEnd[3]{}
\providecommand*\EndOfBibitem{}
\mciteSetBstSublistMode{f}
\mciteSetBstMaxWidthForm{subitem}{(\alph{mcitesubitemcount})}
\mciteSetBstSublistLabelBeginEnd
  {\mcitemaxwidthsubitemform\space}
  {\relax}
  {\relax}

\bibitem[Gouy(1910)]{gouy1910constitution}
Gouy,~M. Sur la constitution de la charge {\'e}lectrique {\`a} la surface d'un {\'e}lectrolyte. \emph{Journal de Physique Théorique et Appliquée} \textbf{1910}, \emph{9}, 457--468\relax
\mciteBstWouldAddEndPuncttrue
\mciteSetBstMidEndSepPunct{\mcitedefaultmidpunct}
{\mcitedefaultendpunct}{\mcitedefaultseppunct}\relax
\EndOfBibitem
\bibitem[Chapman(1913)]{chapman1913li}
Chapman,~D.~L. A contribution to the theory of electrocapillarity. \emph{The London, Edinburgh, and Dublin Philosophical Magazine and Journal of Science} \textbf{1913}, \emph{25}, 475--481\relax
\mciteBstWouldAddEndPuncttrue
\mciteSetBstMidEndSepPunct{\mcitedefaultmidpunct}
{\mcitedefaultendpunct}{\mcitedefaultseppunct}\relax
\EndOfBibitem
\bibitem[Pellenq \latin{et~al.}(2009)Pellenq, Kushima, Shahsavari, Van~Vliet, Buehler, Yip, and Ulm]{pellenq2009realistic}
Pellenq,~R. J.-M.; Kushima,~A.; Shahsavari,~R.; Van~Vliet,~K.~J.; Buehler,~M.~J.; Yip,~S.; Ulm,~F.-J. A realistic molecular model of cement hydrates. \emph{Proceedings of the National Academy of Sciences} \textbf{2009}, \emph{106}, 16102--16107\relax
\mciteBstWouldAddEndPuncttrue
\mciteSetBstMidEndSepPunct{\mcitedefaultmidpunct}
{\mcitedefaultendpunct}{\mcitedefaultseppunct}\relax
\EndOfBibitem
\bibitem[Skibsted \latin{et~al.}(1994)Skibsted, Jakobsen, and Hall]{skibsted1994direct}
Skibsted,~J.; Jakobsen,~H.~J.; Hall,~C. Direct observation of aluminium guest ions in the silicate phases of cement minerals by 27 Al MAS NMR spectroscopy. \emph{Journal of the Chemical Society, Faraday Transactions} \textbf{1994}, \emph{90}, 2095--2098\relax
\mciteBstWouldAddEndPuncttrue
\mciteSetBstMidEndSepPunct{\mcitedefaultmidpunct}
{\mcitedefaultendpunct}{\mcitedefaultseppunct}\relax
\EndOfBibitem
\bibitem[Pellenq \latin{et~al.}(1997)Pellenq, Caillol, and Delville]{pellenq1997electrostatic}
Pellenq,~R.-M.; Caillol,~J.; Delville,~A. Electrostatic attraction between two charged surfaces: A (N, V, T) Monte Carlo simulation. \emph{The Journal of Physical Chemistry B} \textbf{1997}, \emph{101}, 8584--8594\relax
\mciteBstWouldAddEndPuncttrue
\mciteSetBstMidEndSepPunct{\mcitedefaultmidpunct}
{\mcitedefaultendpunct}{\mcitedefaultseppunct}\relax
\EndOfBibitem
\bibitem[Goyal \latin{et~al.}(2021)Goyal, Palaia, Ioannidou, Ulm, Van~Damme, Pellenq, Trizac, and Del~Gado]{goyal2021physics}
Goyal,~A.; Palaia,~I.; Ioannidou,~K.; Ulm,~F.-J.; Van~Damme,~H.; Pellenq,~R. J.-M.; Trizac,~E.; Del~Gado,~E. The physics of cement cohesion. \emph{Science Advances} \textbf{2021}, \emph{7}, eabg5882\relax
\mciteBstWouldAddEndPuncttrue
\mciteSetBstMidEndSepPunct{\mcitedefaultmidpunct}
{\mcitedefaultendpunct}{\mcitedefaultseppunct}\relax
\EndOfBibitem
\bibitem[Chmiola \latin{et~al.}(2006)Chmiola, Yushin, Gogotsi, Portet, Simon, and Taberna]{chmiola2006anomalous}
Chmiola,~J.; Yushin,~G.; Gogotsi,~Y.; Portet,~C.; Simon,~P.; Taberna,~P.-L. Anomalous increase in carbon capacitance at pore sizes less than 1 nanometer. \emph{Science} \textbf{2006}, \emph{313}, 1760--1763\relax
\mciteBstWouldAddEndPuncttrue
\mciteSetBstMidEndSepPunct{\mcitedefaultmidpunct}
{\mcitedefaultendpunct}{\mcitedefaultseppunct}\relax
\EndOfBibitem
\bibitem[Merlet \latin{et~al.}(2012)Merlet, Rotenberg, Madden, Taberna, Simon, Gogotsi, and Salanne]{merlet2012molecular}
Merlet,~C.; Rotenberg,~B.; Madden,~P.~A.; Taberna,~P.-L.; Simon,~P.; Gogotsi,~Y.; Salanne,~M. On the molecular origin of supercapacitance in nanoporous carbon electrodes. \emph{Nature Materials} \textbf{2012}, \emph{11}, 306--310\relax
\mciteBstWouldAddEndPuncttrue
\mciteSetBstMidEndSepPunct{\mcitedefaultmidpunct}
{\mcitedefaultendpunct}{\mcitedefaultseppunct}\relax
\EndOfBibitem
\bibitem[Park and Sposito(2002)Park, and Sposito]{park2002structure}
Park,~S.-H.; Sposito,~G. Structure of water adsorbed on a mica surface. \emph{Physical Review Letters} \textbf{2002}, \emph{89}, 085501\relax
\mciteBstWouldAddEndPuncttrue
\mciteSetBstMidEndSepPunct{\mcitedefaultmidpunct}
{\mcitedefaultendpunct}{\mcitedefaultseppunct}\relax
\EndOfBibitem
\bibitem[Park \latin{et~al.}(2006)Park, Fenter, Nagy, and Sturchio]{park2006hydration}
Park,~C.; Fenter,~P.~A.; Nagy,~K.~L.; Sturchio,~N.~C. Hydration and distribution of ions at the mica-water interface. \emph{Physical Review Letters} \textbf{2006}, \emph{97}, 016101\relax
\mciteBstWouldAddEndPuncttrue
\mciteSetBstMidEndSepPunct{\mcitedefaultmidpunct}
{\mcitedefaultendpunct}{\mcitedefaultseppunct}\relax
\EndOfBibitem
\bibitem[Pashley and Israelachvili(1984)Pashley, and Israelachvili]{pashley1984molecular}
Pashley,~R.~M.; Israelachvili,~J.~N. Molecular layering of water in thin films between mica surfaces and its relation to hydration forces. \emph{Journal of Colloid and Interface Science} \textbf{1984}, \emph{101}, 511--523\relax
\mciteBstWouldAddEndPuncttrue
\mciteSetBstMidEndSepPunct{\mcitedefaultmidpunct}
{\mcitedefaultendpunct}{\mcitedefaultseppunct}\relax
\EndOfBibitem
\bibitem[Plassard \latin{et~al.}(2005)Plassard, Lesniewska, Pochard, and Nonat]{plassard2005nanoscale}
Plassard,~C.; Lesniewska,~E.; Pochard,~I.; Nonat,~A. Nanoscale experimental investigation of particle interactions at the origin of the cohesion of cement. \emph{Langmuir} \textbf{2005}, \emph{21}, 7263--7270\relax
\mciteBstWouldAddEndPuncttrue
\mciteSetBstMidEndSepPunct{\mcitedefaultmidpunct}
{\mcitedefaultendpunct}{\mcitedefaultseppunct}\relax
\EndOfBibitem
\bibitem[L{\"u}tzenkirchen \latin{et~al.}(2012)L{\"u}tzenkirchen, Preo{\v{c}}anin, Kova{\v{c}}evi{\'c}, Tomi{\v{s}}i{\'c}, L{\"o}vgren, and Kallay]{lutzenkirchen2012potentiometric}
L{\"u}tzenkirchen,~J.; Preo{\v{c}}anin,~T.; Kova{\v{c}}evi{\'c},~D.; Tomi{\v{s}}i{\'c},~V.; L{\"o}vgren,~L.; Kallay,~N. Potentiometric titrations as a tool for surface charge determination. \emph{Croatica Chemica Acta} \textbf{2012}, \emph{85}, 391--417\relax
\mciteBstWouldAddEndPuncttrue
\mciteSetBstMidEndSepPunct{\mcitedefaultmidpunct}
{\mcitedefaultendpunct}{\mcitedefaultseppunct}\relax
\EndOfBibitem
\bibitem[Zaera(2012)]{zaera2012probing}
Zaera,~F. Probing liquid/solid interfaces at the molecular level. \emph{Chemical Reviews} \textbf{2012}, \emph{112}, 2920--2986\relax
\mciteBstWouldAddEndPuncttrue
\mciteSetBstMidEndSepPunct{\mcitedefaultmidpunct}
{\mcitedefaultendpunct}{\mcitedefaultseppunct}\relax
\EndOfBibitem
\bibitem[Lee \latin{et~al.}(2021)Lee, Koishi, Bourg, and Fenter]{lee2021ion}
Lee,~S.~S.; Koishi,~A.; Bourg,~I.~C.; Fenter,~P. Ion correlations drive charge overscreening and heterogeneous nucleation at solid--aqueous electrolyte interfaces. \emph{Proceedings of the National Academy of Sciences} \textbf{2021}, \emph{118}, e2105154118\relax
\mciteBstWouldAddEndPuncttrue
\mciteSetBstMidEndSepPunct{\mcitedefaultmidpunct}
{\mcitedefaultendpunct}{\mcitedefaultseppunct}\relax
\EndOfBibitem
\bibitem[Roth and Gillespie(2016)Roth, and Gillespie]{roth2016shells}
Roth,~R.; Gillespie,~D. Shells of charge: a density functional theory for charged hard spheres. \emph{Journal of Physics: Condensed Matter} \textbf{2016}, \emph{28}, 244006\relax
\mciteBstWouldAddEndPuncttrue
\mciteSetBstMidEndSepPunct{\mcitedefaultmidpunct}
{\mcitedefaultendpunct}{\mcitedefaultseppunct}\relax
\EndOfBibitem
\bibitem[Torrie and Valleau(1980)Torrie, and Valleau]{torrie1980electrical}
Torrie,~G.; Valleau,~J. Electrical double layers. I. Monte Carlo study of a uniformly charged surface. \emph{The Journal of Chemical Physics} \textbf{1980}, \emph{73}, 5807--5816\relax
\mciteBstWouldAddEndPuncttrue
\mciteSetBstMidEndSepPunct{\mcitedefaultmidpunct}
{\mcitedefaultendpunct}{\mcitedefaultseppunct}\relax
\EndOfBibitem
\bibitem[Qiao and Aluru(2004)Qiao, and Aluru]{qiao2004charge}
Qiao,~R.; Aluru,~N.~R. Charge inversion and flow reversal in a nanochannel electro-osmotic flow. \emph{Physical Review Letters} \textbf{2004}, \emph{92}, 198301\relax
\mciteBstWouldAddEndPuncttrue
\mciteSetBstMidEndSepPunct{\mcitedefaultmidpunct}
{\mcitedefaultendpunct}{\mcitedefaultseppunct}\relax
\EndOfBibitem
\bibitem[Liu and Eisenberg(2013)Liu, and Eisenberg]{liu2013correlated}
Liu,~J.-L.; Eisenberg,~B. Correlated ions in a calcium channel model: a Poisson--Fermi theory. \emph{The Journal of Physical Chemistry B} \textbf{2013}, \emph{117}, 12051--12058\relax
\mciteBstWouldAddEndPuncttrue
\mciteSetBstMidEndSepPunct{\mcitedefaultmidpunct}
{\mcitedefaultendpunct}{\mcitedefaultseppunct}\relax
\EndOfBibitem
\bibitem[Bazant \latin{et~al.}(2011)Bazant, Storey, and Kornyshev]{bazant2011double}
Bazant,~M.~Z.; Storey,~B.~D.; Kornyshev,~A.~A. Double layer in ionic liquids: Overscreening versus crowding. \emph{Physical Review Letters} \textbf{2011}, \emph{106}, 046102\relax
\mciteBstWouldAddEndPuncttrue
\mciteSetBstMidEndSepPunct{\mcitedefaultmidpunct}
{\mcitedefaultendpunct}{\mcitedefaultseppunct}\relax
\EndOfBibitem
\bibitem[Kierlik and Rosinberg(1991)Kierlik, and Rosinberg]{kierlik1991density}
Kierlik,~E.; Rosinberg,~M. Density-functional theory for inhomogeneous fluids: adsorption of binary mixtures. \emph{Physical Review A} \textbf{1991}, \emph{44}, 5025\relax
\mciteBstWouldAddEndPuncttrue
\mciteSetBstMidEndSepPunct{\mcitedefaultmidpunct}
{\mcitedefaultendpunct}{\mcitedefaultseppunct}\relax
\EndOfBibitem
\bibitem[Gillespie \latin{et~al.}(2002)Gillespie, Nonner, and Eisenberg]{gillespie2002coupling}
Gillespie,~D.; Nonner,~W.; Eisenberg,~R.~S. Coupling Poisson--Nernst--Planck and density functional theory to calculate ion flux. \emph{Journal of Physics: Condensed Matter} \textbf{2002}, \emph{14}, 12129\relax
\mciteBstWouldAddEndPuncttrue
\mciteSetBstMidEndSepPunct{\mcitedefaultmidpunct}
{\mcitedefaultendpunct}{\mcitedefaultseppunct}\relax
\EndOfBibitem
\bibitem[Gupta \latin{et~al.}(2020)Gupta, Rajan, Carter, and Stone]{gupta2020ionic}
Gupta,~A.; Rajan,~A.~G.; Carter,~E.~A.; Stone,~H.~A. Ionic Layering and overcharging in electrical double layers in a Poisson-Boltzmann model. \emph{Physical Review Letters} \textbf{2020}, \emph{125}, 188004\relax
\mciteBstWouldAddEndPuncttrue
\mciteSetBstMidEndSepPunct{\mcitedefaultmidpunct}
{\mcitedefaultendpunct}{\mcitedefaultseppunct}\relax
\EndOfBibitem
\bibitem[de~Souza \latin{et~al.}(2020)de~Souza, Goodwin, McEldrew, Kornyshev, and Bazant]{de2020interfacial}
de~Souza,~J.~P.; Goodwin,~Z.~A.; McEldrew,~M.; Kornyshev,~A.~A.; Bazant,~M.~Z. Interfacial layering in the electric double layer of ionic liquids. \emph{Physical Review Letters} \textbf{2020}, \emph{125}, 116001\relax
\mciteBstWouldAddEndPuncttrue
\mciteSetBstMidEndSepPunct{\mcitedefaultmidpunct}
{\mcitedefaultendpunct}{\mcitedefaultseppunct}\relax
\EndOfBibitem
\bibitem[Chapman \latin{et~al.}(1989)Chapman, Gubbins, Jackson, and Radosz]{chapman1989saft}
Chapman,~W.~G.; Gubbins,~K.~E.; Jackson,~G.; Radosz,~M. SAFT: Equation-of-state solution model for associating fluids. \emph{Fluid Phase Equilibria} \textbf{1989}, \emph{52}, 31--38\relax
\mciteBstWouldAddEndPuncttrue
\mciteSetBstMidEndSepPunct{\mcitedefaultmidpunct}
{\mcitedefaultendpunct}{\mcitedefaultseppunct}\relax
\EndOfBibitem
\bibitem[Hansen and McDonald(2013)Hansen, and McDonald]{hansen2013theory}
Hansen,~J.-P.; McDonald,~I.~R. \emph{Theory of simple liquids: with applications to soft matter}; Academic press, 2013\relax
\mciteBstWouldAddEndPuncttrue
\mciteSetBstMidEndSepPunct{\mcitedefaultmidpunct}
{\mcitedefaultendpunct}{\mcitedefaultseppunct}\relax
\EndOfBibitem
\bibitem[Tarazona \latin{et~al.}(2008)Tarazona, Cuesta, and Mart{\'\i}nez-Rat{\'o}n]{tarazona2008density}
Tarazona,~P.; Cuesta,~J.~A.; Mart{\'\i}nez-Rat{\'o}n,~Y. \emph{Theory and Simulation of Hard-Sphere Fluids and Related Systems}; Springer, 2008; pp 247--341\relax
\mciteBstWouldAddEndPuncttrue
\mciteSetBstMidEndSepPunct{\mcitedefaultmidpunct}
{\mcitedefaultendpunct}{\mcitedefaultseppunct}\relax
\EndOfBibitem
\bibitem[Siderius and Gelb(2011)Siderius, and Gelb]{siderius2011extension}
Siderius,~D.~W.; Gelb,~L.~D. Extension of the Steele 10-4-3 potential for adsorption calculations in cylindrical, spherical, and other pore geometries. \emph{The Journal of Chemical Physics} \textbf{2011}, \emph{135}, 084703\relax
\mciteBstWouldAddEndPuncttrue
\mciteSetBstMidEndSepPunct{\mcitedefaultmidpunct}
{\mcitedefaultendpunct}{\mcitedefaultseppunct}\relax
\EndOfBibitem
\bibitem[Steele(1973)]{steele1973physical}
Steele,~W.~A. The physical interaction of gases with crystalline solids: I. Gas-solid energies and properties of isolated adsorbed atoms. \emph{Surface Science} \textbf{1973}, \emph{36}, 317--352\relax
\mciteBstWouldAddEndPuncttrue
\mciteSetBstMidEndSepPunct{\mcitedefaultmidpunct}
{\mcitedefaultendpunct}{\mcitedefaultseppunct}\relax
\EndOfBibitem
\bibitem[Balbuena and Gubbins(1993)Balbuena, and Gubbins]{balbuena1993theoretical}
Balbuena,~P.~B.; Gubbins,~K.~E. Theoretical interpretation of adsorption behavior of simple fluids in slit pores. \emph{Langmuir} \textbf{1993}, \emph{9}, 1801--1814\relax
\mciteBstWouldAddEndPuncttrue
\mciteSetBstMidEndSepPunct{\mcitedefaultmidpunct}
{\mcitedefaultendpunct}{\mcitedefaultseppunct}\relax
\EndOfBibitem
\bibitem[Tarazona(2000)]{tarazona2000density}
Tarazona,~P. Density functional for hard sphere crystals: A fundamental measure approach. \emph{Physical Review Letters} \textbf{2000}, \emph{84}, 694\relax
\mciteBstWouldAddEndPuncttrue
\mciteSetBstMidEndSepPunct{\mcitedefaultmidpunct}
{\mcitedefaultendpunct}{\mcitedefaultseppunct}\relax
\EndOfBibitem
\bibitem[Rosenfeld(1989)]{rosenfeld1989free}
Rosenfeld,~Y. Free-energy model for the inhomogeneous hard-sphere fluid mixture and density-functional theory of freezing. \emph{Physical Review Letters} \textbf{1989}, \emph{63}, 980\relax
\mciteBstWouldAddEndPuncttrue
\mciteSetBstMidEndSepPunct{\mcitedefaultmidpunct}
{\mcitedefaultendpunct}{\mcitedefaultseppunct}\relax
\EndOfBibitem
\bibitem[Roth(2010)]{roth2010fundamental}
Roth,~R. Fundamental measure theory for hard-sphere mixtures: a review. \emph{Journal of Physics: Condensed Matter} \textbf{2010}, \emph{22}, 063102\relax
\mciteBstWouldAddEndPuncttrue
\mciteSetBstMidEndSepPunct{\mcitedefaultmidpunct}
{\mcitedefaultendpunct}{\mcitedefaultseppunct}\relax
\EndOfBibitem
\bibitem[Barker and Henderson(1976)Barker, and Henderson]{barker1976liquid}
Barker,~J.~A.; Henderson,~D. What is" liquid"? Understanding the states of matter. \emph{Reviews of Modern Physics} \textbf{1976}, \emph{48}, 587\relax
\mciteBstWouldAddEndPuncttrue
\mciteSetBstMidEndSepPunct{\mcitedefaultmidpunct}
{\mcitedefaultendpunct}{\mcitedefaultseppunct}\relax
\EndOfBibitem
\bibitem[Gil-Villegas \latin{et~al.}(1997)Gil-Villegas, Galindo, Whitehead, Mills, Jackson, and Burgess]{gil1997statistical}
Gil-Villegas,~A.; Galindo,~A.; Whitehead,~P.~J.; Mills,~S.~J.; Jackson,~G.; Burgess,~A.~N. Statistical associating fluid theory for chain molecules with attractive potentials of variable range. \emph{The Journal of Chemical Physics} \textbf{1997}, \emph{106}, 4168--4186\relax
\mciteBstWouldAddEndPuncttrue
\mciteSetBstMidEndSepPunct{\mcitedefaultmidpunct}
{\mcitedefaultendpunct}{\mcitedefaultseppunct}\relax
\EndOfBibitem
\bibitem[Wertheim(1984)]{wertheim1984fluids_I}
Wertheim,~M. Fluids with highly directional attractive forces. I. Statistical thermodynamics. \emph{Journal of Statistical Physics} \textbf{1984}, \emph{35}, 19--34\relax
\mciteBstWouldAddEndPuncttrue
\mciteSetBstMidEndSepPunct{\mcitedefaultmidpunct}
{\mcitedefaultendpunct}{\mcitedefaultseppunct}\relax
\EndOfBibitem
\bibitem[Wertheim(1984)]{wertheim1984fluids_II}
Wertheim,~M.~S. Fluids with highly directional attractive forces. II. Thermodynamic perturbation theory and integral equations. \emph{Journal of Statistical Physics} \textbf{1984}, \emph{35}, 35--47\relax
\mciteBstWouldAddEndPuncttrue
\mciteSetBstMidEndSepPunct{\mcitedefaultmidpunct}
{\mcitedefaultendpunct}{\mcitedefaultseppunct}\relax
\EndOfBibitem
\bibitem[Wertheim(1986)]{wertheim1986fluids_III}
Wertheim,~M. Fluids with highly directional attractive forces. III. Multiple attraction sites. \emph{Journal of Statistical Physics} \textbf{1986}, \emph{42}, 459--476\relax
\mciteBstWouldAddEndPuncttrue
\mciteSetBstMidEndSepPunct{\mcitedefaultmidpunct}
{\mcitedefaultendpunct}{\mcitedefaultseppunct}\relax
\EndOfBibitem
\bibitem[Wertheim(1986)]{wertheim1986fluids_IV}
Wertheim,~M. Fluids with highly directional attractive forces. IV. Equilibrium polymerization. \emph{Journal of Statistical Physics} \textbf{1986}, \emph{42}, 477--492\relax
\mciteBstWouldAddEndPuncttrue
\mciteSetBstMidEndSepPunct{\mcitedefaultmidpunct}
{\mcitedefaultendpunct}{\mcitedefaultseppunct}\relax
\EndOfBibitem
\bibitem[Yu and Wu(2002)Yu, and Wu]{yu2002fundamental}
Yu,~Y.-X.; Wu,~J. A fundamental-measure theory for inhomogeneous associating fluids. \emph{The Journal of Chemical Physics} \textbf{2002}, \emph{116}, 7094--7103\relax
\mciteBstWouldAddEndPuncttrue
\mciteSetBstMidEndSepPunct{\mcitedefaultmidpunct}
{\mcitedefaultendpunct}{\mcitedefaultseppunct}\relax
\EndOfBibitem
\bibitem[Camacho~Vergara \latin{et~al.}(2020)Camacho~Vergara, Kontogeorgis, and Liang]{camacho2020new}
Camacho~Vergara,~E.~L.; Kontogeorgis,~G.~M.; Liang,~X. A new study of associating inhomogeneous fluids with classical density functional theory. \emph{Molecular Physics} \textbf{2020}, \emph{118}, e1725668\relax
\mciteBstWouldAddEndPuncttrue
\mciteSetBstMidEndSepPunct{\mcitedefaultmidpunct}
{\mcitedefaultendpunct}{\mcitedefaultseppunct}\relax
\EndOfBibitem
\bibitem[Segura and Chapman(1995)Segura, and Chapman]{segura1995associating}
Segura,~C.~J.; Chapman,~W.~G. Associating fluids with four bonding sites against solid surfaces: Monte Carlo simulations. \emph{Molecular Physics} \textbf{1995}, \emph{86}, 415--442\relax
\mciteBstWouldAddEndPuncttrue
\mciteSetBstMidEndSepPunct{\mcitedefaultmidpunct}
{\mcitedefaultendpunct}{\mcitedefaultseppunct}\relax
\EndOfBibitem
\bibitem[Segura \latin{et~al.}(1997)Segura, Chapman, and Shukla]{segura1997associating}
Segura,~C.~J.; Chapman,~W.~G.; Shukla,~K.~P. Associating fluids with four bonding sites against a hard wall: density functional theory. \emph{Molecular Physics} \textbf{1997}, \emph{90}, 759--772\relax
\mciteBstWouldAddEndPuncttrue
\mciteSetBstMidEndSepPunct{\mcitedefaultmidpunct}
{\mcitedefaultendpunct}{\mcitedefaultseppunct}\relax
\EndOfBibitem
\bibitem[Gloor \latin{et~al.}(2007)Gloor, Jackson, Blas, Del~Rio, and De~Miguel]{gloor2007prediction}
Gloor,~G.~J.; Jackson,~G.; Blas,~F.; Del~Rio,~E.~M.; De~Miguel,~E. Prediction of the vapor- liquid interfacial tension of nonassociating and associating fluids with the SAFT-VR density functional theory. \emph{The Journal of Physical Chemistry C} \textbf{2007}, \emph{111}, 15513--15522\relax
\mciteBstWouldAddEndPuncttrue
\mciteSetBstMidEndSepPunct{\mcitedefaultmidpunct}
{\mcitedefaultendpunct}{\mcitedefaultseppunct}\relax
\EndOfBibitem
\bibitem[Mac~Dowell \latin{et~al.}(2010)Mac~Dowell, Llovell, Adjiman, Jackson, and Galindo]{mac2010modeling}
Mac~Dowell,~N.; Llovell,~F.; Adjiman,~C.; Jackson,~G.; Galindo,~A. Modeling the fluid phase behavior of carbon dioxide in aqueous solutions of monoethanolamine using transferable parameters with the SAFT-VR approach. \emph{Industrial \& Engineering Chemistry Research} \textbf{2010}, \emph{49}, 1883--1899\relax
\mciteBstWouldAddEndPuncttrue
\mciteSetBstMidEndSepPunct{\mcitedefaultmidpunct}
{\mcitedefaultendpunct}{\mcitedefaultseppunct}\relax
\EndOfBibitem
\bibitem[Pereira \latin{et~al.}(2011)Pereira, Keskes, Galindo, Jackson, and Adjiman]{pereira2011integrated}
Pereira,~F.~E.; Keskes,~E.; Galindo,~A.; Jackson,~G.; Adjiman,~C.~S. Integrated solvent and process design using a SAFT-VR thermodynamic description: High-pressure separation of carbon dioxide and methane. \emph{Computers \& Chemical Engineering} \textbf{2011}, \emph{35}, 474--491\relax
\mciteBstWouldAddEndPuncttrue
\mciteSetBstMidEndSepPunct{\mcitedefaultmidpunct}
{\mcitedefaultendpunct}{\mcitedefaultseppunct}\relax
\EndOfBibitem
\bibitem[Blum(1975)]{blum1975mean}
Blum,~L. Mean spherical model for asymmetric electrolytes: I. Method of solution. \emph{Molecular Physics} \textbf{1975}, \emph{30}, 1529--1535\relax
\mciteBstWouldAddEndPuncttrue
\mciteSetBstMidEndSepPunct{\mcitedefaultmidpunct}
{\mcitedefaultendpunct}{\mcitedefaultseppunct}\relax
\EndOfBibitem
\bibitem[Blum and Wei(1987)Blum, and Wei]{blum1987analytical}
Blum,~L.; Wei,~d. Analytical solution of the mean spherical approximation for an arbitrary mixture of ions in a dipolar solvent. \emph{The Journal of Chemical Physics} \textbf{1987}, \emph{87}, 555--565\relax
\mciteBstWouldAddEndPuncttrue
\mciteSetBstMidEndSepPunct{\mcitedefaultmidpunct}
{\mcitedefaultendpunct}{\mcitedefaultseppunct}\relax
\EndOfBibitem
\bibitem[Voukadinova \latin{et~al.}(2018)Voukadinova, Valisk{\'o}, and Gillespie]{voukadinova2018assessing}
Voukadinova,~A.; Valisk{\'o},~M.; Gillespie,~D. Assessing the accuracy of three classical density functional theories of the electrical double layer. \emph{Physical Review E} \textbf{2018}, \emph{98}, 012116\relax
\mciteBstWouldAddEndPuncttrue
\mciteSetBstMidEndSepPunct{\mcitedefaultmidpunct}
{\mcitedefaultendpunct}{\mcitedefaultseppunct}\relax
\EndOfBibitem
\bibitem[Rosenfeld(1993)]{rosenfeld1993free}
Rosenfeld,~Y. Free energy model for inhomogeneous fluid mixtures: Yukawa-charged hard spheres, general interactions, and plasmas. \emph{The Journal of Chemical Physics} \textbf{1993}, \emph{98}, 8126--8148\relax
\mciteBstWouldAddEndPuncttrue
\mciteSetBstMidEndSepPunct{\mcitedefaultmidpunct}
{\mcitedefaultendpunct}{\mcitedefaultseppunct}\relax
\EndOfBibitem
\bibitem[Jiang and Gillespie(2021)Jiang, and Gillespie]{jiang2021revisiting}
Jiang,~J.; Gillespie,~D. Revisiting the charged shell model: A density functional theory for electrolytes. \emph{Journal of Chemical Theory and Computation} \textbf{2021}, \emph{17}, 2409--2416\relax
\mciteBstWouldAddEndPuncttrue
\mciteSetBstMidEndSepPunct{\mcitedefaultmidpunct}
{\mcitedefaultendpunct}{\mcitedefaultseppunct}\relax
\EndOfBibitem
\bibitem[de~Souza \latin{et~al.}(2022)de~Souza, Kornyshev, and Bazant]{de2022polar}
de~Souza,~J.~P.; Kornyshev,~A.~A.; Bazant,~M.~Z. Polar liquids at charged interfaces: A dipolar shell theory. \emph{The Journal of Chemical Physics} \textbf{2022}, \emph{156}, 244705\relax
\mciteBstWouldAddEndPuncttrue
\mciteSetBstMidEndSepPunct{\mcitedefaultmidpunct}
{\mcitedefaultendpunct}{\mcitedefaultseppunct}\relax
\EndOfBibitem
\bibitem[H{\"a}rtel(2013)]{hartel2013density}
H{\"a}rtel,~A. Density functional theory of hard colloidal particles: From bulk to interfaces. Ph.D.\ thesis, D{\"u}sseldorf, Heinrich-Heine-Universit{\"a}t, Diss., 2013, 2013\relax
\mciteBstWouldAddEndPuncttrue
\mciteSetBstMidEndSepPunct{\mcitedefaultmidpunct}
{\mcitedefaultendpunct}{\mcitedefaultseppunct}\relax
\EndOfBibitem
\bibitem[Barker and Henderson(1967)Barker, and Henderson]{barker1967perturbation}
Barker,~J.~A.; Henderson,~D. Perturbation theory and equation of state for fluids: the square-well potential. \emph{The Journal of Chemical Physics} \textbf{1967}, \emph{47}, 2856--2861\relax
\mciteBstWouldAddEndPuncttrue
\mciteSetBstMidEndSepPunct{\mcitedefaultmidpunct}
{\mcitedefaultendpunct}{\mcitedefaultseppunct}\relax
\EndOfBibitem
\bibitem[Barker and Henderson(1967)Barker, and Henderson]{barker1967perturbationII}
Barker,~J.~A.; Henderson,~D. Perturbation theory and equation of state for fluids. II. A successful theory of liquids. \emph{The Journal of Chemical Physics} \textbf{1967}, \emph{47}, 4714--4721\relax
\mciteBstWouldAddEndPuncttrue
\mciteSetBstMidEndSepPunct{\mcitedefaultmidpunct}
{\mcitedefaultendpunct}{\mcitedefaultseppunct}\relax
\EndOfBibitem
\bibitem[Weeks \latin{et~al.}(1971)Weeks, Chandler, and Andersen]{weeks1971role}
Weeks,~J.~D.; Chandler,~D.; Andersen,~H.~C. Role of repulsive forces in determining the equilibrium structure of simple liquids. \emph{The Journal of Chemical Physics} \textbf{1971}, \emph{54}, 5237--5247\relax
\mciteBstWouldAddEndPuncttrue
\mciteSetBstMidEndSepPunct{\mcitedefaultmidpunct}
{\mcitedefaultendpunct}{\mcitedefaultseppunct}\relax
\EndOfBibitem
\bibitem[Percus and Yevick(1958)Percus, and Yevick]{percus1958analysis}
Percus,~J.~K.; Yevick,~G.~J. Analysis of classical statistical mechanics by means of collective coordinates. \emph{The Physical Review} \textbf{1958}, \emph{110}, 1\relax
\mciteBstWouldAddEndPuncttrue
\mciteSetBstMidEndSepPunct{\mcitedefaultmidpunct}
{\mcitedefaultendpunct}{\mcitedefaultseppunct}\relax
\EndOfBibitem
\bibitem[Gross and Sadowski(2001)Gross, and Sadowski]{gross2001perturbed}
Gross,~J.; Sadowski,~G. Perturbed-chain SAFT: An equation of state based on a perturbation theory for chain molecules. \emph{Industrial \& Engineering Chemistry Research} \textbf{2001}, \emph{40}, 1244--1260\relax
\mciteBstWouldAddEndPuncttrue
\mciteSetBstMidEndSepPunct{\mcitedefaultmidpunct}
{\mcitedefaultendpunct}{\mcitedefaultseppunct}\relax
\EndOfBibitem
\bibitem[Llovell \latin{et~al.}(2010)Llovell, Galindo, Blas, and Jackson]{llovell2010classical}
Llovell,~F.; Galindo,~A.; Blas,~F.~J.; Jackson,~G. Classical density functional theory for the prediction of the surface tension and interfacial properties of fluids mixtures of chain molecules based on the statistical associating fluid theory for potentials of variable range. \emph{The Journal of Chemical Physics} \textbf{2010}, \emph{133}, 024704\relax
\mciteBstWouldAddEndPuncttrue
\mciteSetBstMidEndSepPunct{\mcitedefaultmidpunct}
{\mcitedefaultendpunct}{\mcitedefaultseppunct}\relax
\EndOfBibitem
\bibitem[Carnahan and Starling(1969)Carnahan, and Starling]{carnahan1969equation}
Carnahan,~N.~F.; Starling,~K.~E. Equation of state for nonattracting rigid spheres. \emph{The Journal of Chemical Physics} \textbf{1969}, \emph{51}, 635--636\relax
\mciteBstWouldAddEndPuncttrue
\mciteSetBstMidEndSepPunct{\mcitedefaultmidpunct}
{\mcitedefaultendpunct}{\mcitedefaultseppunct}\relax
\EndOfBibitem
\bibitem[Boubl{\'\i}k(1970)]{boublik1970hard}
Boubl{\'\i}k,~T. Hard-sphere equation of state. \emph{The Journal of chemical physics} \textbf{1970}, \emph{53}, 471--472\relax
\mciteBstWouldAddEndPuncttrue
\mciteSetBstMidEndSepPunct{\mcitedefaultmidpunct}
{\mcitedefaultendpunct}{\mcitedefaultseppunct}\relax
\EndOfBibitem
\bibitem[Mansoori \latin{et~al.}(1971)Mansoori, Carnahan, Starling, and Leland~Jr]{mansoori1971equilibrium}
Mansoori,~G.; Carnahan,~N.~F.; Starling,~K.; Leland~Jr,~T. Equilibrium thermodynamic properties of the mixture of hard spheres. \emph{The Journal of Chemical Physics} \textbf{1971}, \emph{54}, 1523--1525\relax
\mciteBstWouldAddEndPuncttrue
\mciteSetBstMidEndSepPunct{\mcitedefaultmidpunct}
{\mcitedefaultendpunct}{\mcitedefaultseppunct}\relax
\EndOfBibitem
\bibitem[Kirkwood(1939)]{kirkwood1939dielectric}
Kirkwood,~J.~G. The dielectric polarization of polar liquids. \emph{The Journal of Chemical Physics} \textbf{1939}, \emph{7}, 911--919\relax
\mciteBstWouldAddEndPuncttrue
\mciteSetBstMidEndSepPunct{\mcitedefaultmidpunct}
{\mcitedefaultendpunct}{\mcitedefaultseppunct}\relax
\EndOfBibitem
\bibitem[Bonthuis \latin{et~al.}(2011)Bonthuis, Gekle, and Netz]{bonthuis2011dielectric}
Bonthuis,~D.~J.; Gekle,~S.; Netz,~R.~R. Dielectric profile of interfacial water and its effect on double-layer capacitance. \emph{Physical Review Letters} \textbf{2011}, \emph{107}, 166102\relax
\mciteBstWouldAddEndPuncttrue
\mciteSetBstMidEndSepPunct{\mcitedefaultmidpunct}
{\mcitedefaultendpunct}{\mcitedefaultseppunct}\relax
\EndOfBibitem
\bibitem[Blum(1978)]{blum1978solution}
Blum,~L. Solution of the mean spherical approximation for hard ions and dipoles of arbitrary size. \emph{Journal of Statistical Physics} \textbf{1978}, \emph{18}, 451--474\relax
\mciteBstWouldAddEndPuncttrue
\mciteSetBstMidEndSepPunct{\mcitedefaultmidpunct}
{\mcitedefaultendpunct}{\mcitedefaultseppunct}\relax
\EndOfBibitem
\bibitem[Simonin(2020)]{simonin2020solution}
Simonin,~J.-P. On the solution of the mean-spherical approximation (MSA) for ions in a dipolar solvent in the general case. \emph{AIP Advances} \textbf{2020}, \emph{10}, 114502\relax
\mciteBstWouldAddEndPuncttrue
\mciteSetBstMidEndSepPunct{\mcitedefaultmidpunct}
{\mcitedefaultendpunct}{\mcitedefaultseppunct}\relax
\EndOfBibitem
\bibitem[Simonin and H{\o}ye(2021)Simonin, and H{\o}ye]{simonin2021full}
Simonin,~J.-P.; H{\o}ye,~J.~S. Full solution to the mean spherical approximation (MSA) for an arbitrary mixture of ions in a dipolar solvent. \emph{The Journal of Chemical Physics} \textbf{2021}, \emph{155}\relax
\mciteBstWouldAddEndPuncttrue
\mciteSetBstMidEndSepPunct{\mcitedefaultmidpunct}
{\mcitedefaultendpunct}{\mcitedefaultseppunct}\relax
\EndOfBibitem
\bibitem[Holovko and Protsykevich(2018)Holovko, and Protsykevich]{holovko2018application}
Holovko,~M.; Protsykevich,~I. On the application of the associative mean spherical approximation to the ion-dipole model for electrolyte solutions. \emph{Journal of Molecular Liquids} \textbf{2018}, \emph{270}, 46--51\relax
\mciteBstWouldAddEndPuncttrue
\mciteSetBstMidEndSepPunct{\mcitedefaultmidpunct}
{\mcitedefaultendpunct}{\mcitedefaultseppunct}\relax
\EndOfBibitem
\bibitem[Huang and Radosz(1990)Huang, and Radosz]{huang1990equation}
Huang,~S.~H.; Radosz,~M. Equation of state for small, large, polydisperse, and associating molecules. \emph{Industrial \& Engineering Chemistry Research} \textbf{1990}, \emph{29}, 2284--2294\relax
\mciteBstWouldAddEndPuncttrue
\mciteSetBstMidEndSepPunct{\mcitedefaultmidpunct}
{\mcitedefaultendpunct}{\mcitedefaultseppunct}\relax
\EndOfBibitem
\bibitem[Stopper \latin{et~al.}(2018)Stopper, Hirschmann, Oettel, and Roth]{stopper2018bulk}
Stopper,~D.; Hirschmann,~F.; Oettel,~M.; Roth,~R. Bulk structural information from density functionals for patchy particles. \emph{The Journal of Chemical Physics} \textbf{2018}, \emph{149}, 224503\relax
\mciteBstWouldAddEndPuncttrue
\mciteSetBstMidEndSepPunct{\mcitedefaultmidpunct}
{\mcitedefaultendpunct}{\mcitedefaultseppunct}\relax
\EndOfBibitem
\bibitem[Reed and Gubbins(1973)Reed, and Gubbins]{reed1973applied}
Reed,~T.~M.; Gubbins,~K.~E. \emph{Applied statistical mechanics}; McGraw-Hill, 1973\relax
\mciteBstWouldAddEndPuncttrue
\mciteSetBstMidEndSepPunct{\mcitedefaultmidpunct}
{\mcitedefaultendpunct}{\mcitedefaultseppunct}\relax
\EndOfBibitem
\bibitem[Michelsen and Hendriks(2001)Michelsen, and Hendriks]{michelsen2001physical}
Michelsen,~M.~L.; Hendriks,~E.~M. Physical properties from association models. \emph{Fluid Phase Equilibria} \textbf{2001}, \emph{180}, 165--174\relax
\mciteBstWouldAddEndPuncttrue
\mciteSetBstMidEndSepPunct{\mcitedefaultmidpunct}
{\mcitedefaultendpunct}{\mcitedefaultseppunct}\relax
\EndOfBibitem
\bibitem[Loche \latin{et~al.}(2020)Loche, Ayaz, Wolde-Kidan, Schlaich, and Netz]{loche2020universal}
Loche,~P.; Ayaz,~C.; Wolde-Kidan,~A.; Schlaich,~A.; Netz,~R.~R. Universal and nonuniversal aspects of electrostatics in aqueous nanoconfinement. \emph{The Journal of Physical Chemistry B} \textbf{2020}, \emph{124}, 4365--4371\relax
\mciteBstWouldAddEndPuncttrue
\mciteSetBstMidEndSepPunct{\mcitedefaultmidpunct}
{\mcitedefaultendpunct}{\mcitedefaultseppunct}\relax
\EndOfBibitem
\bibitem[Cheng \latin{et~al.}(2001)Cheng, Fenter, Nagy, Schlegel, and Sturchio]{cheng2001molecular}
Cheng,~L.; Fenter,~P.; Nagy,~K.; Schlegel,~M.; Sturchio,~N. Molecular-scale density oscillations in water adjacent to a mica surface. \emph{Physical Review Letters} \textbf{2001}, \emph{87}, 156103\relax
\mciteBstWouldAddEndPuncttrue
\mciteSetBstMidEndSepPunct{\mcitedefaultmidpunct}
{\mcitedefaultendpunct}{\mcitedefaultseppunct}\relax
\EndOfBibitem
\bibitem[Jackson and West(1931)Jackson, and West]{jackson1931crystal}
Jackson,~W.; West,~J. The Crystal Structure of Muscovite—KAl$_2$ (AlSi$_3$) O$_{10}$ (OH)$_2$. \emph{Zeitschrift f{\"u}r Kristallographie-Crystalline Materials} \textbf{1931}, \emph{76}, 211--227\relax
\mciteBstWouldAddEndPuncttrue
\mciteSetBstMidEndSepPunct{\mcitedefaultmidpunct}
{\mcitedefaultendpunct}{\mcitedefaultseppunct}\relax
\EndOfBibitem
\bibitem[Martin-Jimenez \latin{et~al.}(2016)Martin-Jimenez, Chacon, Tarazona, and Garcia]{martin2016atomically}
Martin-Jimenez,~D.; Chacon,~E.; Tarazona,~P.; Garcia,~R. Atomically resolved three-dimensional structures of electrolyte aqueous solutions near a solid surface. \emph{Nature Communications} \textbf{2016}, \emph{7}, 12164\relax
\mciteBstWouldAddEndPuncttrue
\mciteSetBstMidEndSepPunct{\mcitedefaultmidpunct}
{\mcitedefaultendpunct}{\mcitedefaultseppunct}\relax
\EndOfBibitem
\bibitem[Eriksen \latin{et~al.}(2016)Eriksen, Lazarou, Galindo, Jackson, Adjiman, and Haslam]{eriksen2016development}
Eriksen,~D.~K.; Lazarou,~G.; Galindo,~A.; Jackson,~G.; Adjiman,~C.~S.; Haslam,~A.~J. Development of intermolecular potential models for electrolyte solutions using an electrolyte SAFT-VR Mie equation of state. \emph{Molecular Physics} \textbf{2016}, \emph{114}, 2724--2749\relax
\mciteBstWouldAddEndPuncttrue
\mciteSetBstMidEndSepPunct{\mcitedefaultmidpunct}
{\mcitedefaultendpunct}{\mcitedefaultseppunct}\relax
\EndOfBibitem
\bibitem[Kirchner \latin{et~al.}(2013)Kirchner, Kirchner, Ivani{\v{s}}t{\v{s}}ev, and Fedorov]{kirchner2013electrical}
Kirchner,~K.; Kirchner,~T.; Ivani{\v{s}}t{\v{s}}ev,~V.; Fedorov,~M.~V. Electrical double layer in ionic liquids: Structural transitions from multilayer to monolayer structure at the interface. \emph{Electrochimica Acta} \textbf{2013}, \emph{110}, 762--771\relax
\mciteBstWouldAddEndPuncttrue
\mciteSetBstMidEndSepPunct{\mcitedefaultmidpunct}
{\mcitedefaultendpunct}{\mcitedefaultseppunct}\relax
\EndOfBibitem
\bibitem[Gil-Villegas \latin{et~al.}(2001)Gil-Villegas, Galindo, and Jackson]{gil2001statistical}
Gil-Villegas,~A.; Galindo,~A.; Jackson,~G. A statistical associating fluid theory for electrolyte solutions (SAFT-VRE). \emph{Molecular Physics} \textbf{2001}, \emph{99}, 531--546\relax
\mciteBstWouldAddEndPuncttrue
\mciteSetBstMidEndSepPunct{\mcitedefaultmidpunct}
{\mcitedefaultendpunct}{\mcitedefaultseppunct}\relax
\EndOfBibitem
\bibitem[Bourg \latin{et~al.}(2017)Bourg, Lee, Fenter, and Tournassat]{bourg2017stern}
Bourg,~I.~C.; Lee,~S.~S.; Fenter,~P.; Tournassat,~C. Stern layer structure and energetics at mica--water interfaces. \emph{The Journal of Physical Chemistry C} \textbf{2017}, \emph{121}, 9402--9412\relax
\mciteBstWouldAddEndPuncttrue
\mciteSetBstMidEndSepPunct{\mcitedefaultmidpunct}
{\mcitedefaultendpunct}{\mcitedefaultseppunct}\relax
\EndOfBibitem
\bibitem[Sakuma \latin{et~al.}(2011)Sakuma, Kondo, Nakao, Shiraki, and Kawamura]{sakuma2011structure}
Sakuma,~H.; Kondo,~T.; Nakao,~H.; Shiraki,~K.; Kawamura,~K. Structure of hydrated sodium ions and water molecules adsorbed on the mica/water interface. \emph{The Journal of Physical Chemistry C} \textbf{2011}, \emph{115}, 15959--15964\relax
\mciteBstWouldAddEndPuncttrue
\mciteSetBstMidEndSepPunct{\mcitedefaultmidpunct}
{\mcitedefaultendpunct}{\mcitedefaultseppunct}\relax
\EndOfBibitem
\bibitem[Diao \latin{et~al.}(2019)Diao, Greenwood, Wang, Nam, and Espinosa-Marzal]{diao2019slippery}
Diao,~Y.; Greenwood,~G.; Wang,~M.~C.; Nam,~S.; Espinosa-Marzal,~R.~M. Slippery and sticky graphene in water. \emph{ACS Nano} \textbf{2019}, \emph{13}, 2072--2082\relax
\mciteBstWouldAddEndPuncttrue
\mciteSetBstMidEndSepPunct{\mcitedefaultmidpunct}
{\mcitedefaultendpunct}{\mcitedefaultseppunct}\relax
\EndOfBibitem
\bibitem[Abdolhosseini~Qomi \latin{et~al.}(2014)Abdolhosseini~Qomi, Krakowiak, Bauchy, Stewart, Shahsavari, Jagannathan, Brommer, Baronnet, Buehler, Yip, \latin{et~al.} others]{abdolhosseini2014combinatorial}
Abdolhosseini~Qomi,~M.; Krakowiak,~K.; Bauchy,~M.; Stewart,~K.; Shahsavari,~R.; Jagannathan,~D.; Brommer,~D.~B.; Baronnet,~A.; Buehler,~M.~J.; Yip,~S. \latin{et~al.}  Combinatorial molecular optimization of cement hydrates. \emph{Nature Communications} \textbf{2014}, \emph{5}, 1--10\relax
\mciteBstWouldAddEndPuncttrue
\mciteSetBstMidEndSepPunct{\mcitedefaultmidpunct}
{\mcitedefaultendpunct}{\mcitedefaultseppunct}\relax
\EndOfBibitem
\bibitem[Moreira and Netz(2000)Moreira, and Netz]{moreira2000strong}
Moreira,~A.; Netz,~R. Strong-coupling theory for counter-ion distributions. \emph{Europhysics Letters} \textbf{2000}, \emph{52}, 705\relax
\mciteBstWouldAddEndPuncttrue
\mciteSetBstMidEndSepPunct{\mcitedefaultmidpunct}
{\mcitedefaultendpunct}{\mcitedefaultseppunct}\relax
\EndOfBibitem
\bibitem[Naji \latin{et~al.}(2005)Naji, Jungblut, Moreira, and Netz]{naji2005electrostatic}
Naji,~A.; Jungblut,~S.; Moreira,~A.~G.; Netz,~R.~R. Electrostatic interactions in strongly coupled soft matter. \emph{Physica A: Statistical Mechanics and its Applications} \textbf{2005}, \emph{352}, 131--170\relax
\mciteBstWouldAddEndPuncttrue
\mciteSetBstMidEndSepPunct{\mcitedefaultmidpunct}
{\mcitedefaultendpunct}{\mcitedefaultseppunct}\relax
\EndOfBibitem
\bibitem[{\v{S}}amaj \latin{et~al.}(2020){\v{S}}amaj, Trulsson, and Trizac]{vsamaj2020strong}
{\v{S}}amaj,~L.; Trulsson,~M.; Trizac,~E. Strong-coupling theory of counterions with hard cores between symmetrically charged walls. \emph{Physical Review E} \textbf{2020}, \emph{102}, 042604\relax
\mciteBstWouldAddEndPuncttrue
\mciteSetBstMidEndSepPunct{\mcitedefaultmidpunct}
{\mcitedefaultendpunct}{\mcitedefaultseppunct}\relax
\EndOfBibitem
\bibitem[Gonella \latin{et~al.}(2021)Gonella, Backus, Nagata, Bonthuis, Loche, Schlaich, Netz, K{\"u}hnle, McCrum, Koper, \latin{et~al.} others]{gonella2021water}
Gonella,~G.; Backus,~E.~H.; Nagata,~Y.; Bonthuis,~D.~J.; Loche,~P.; Schlaich,~A.; Netz,~R.~R.; K{\"u}hnle,~A.; McCrum,~I.~T.; Koper,~M.~T. \latin{et~al.}  Water at charged interfaces. \emph{Nature Reviews Chemistry} \textbf{2021}, \emph{5}, 466--485\relax
\mciteBstWouldAddEndPuncttrue
\mciteSetBstMidEndSepPunct{\mcitedefaultmidpunct}
{\mcitedefaultendpunct}{\mcitedefaultseppunct}\relax
\EndOfBibitem
\bibitem[Tielrooij \latin{et~al.}(2009)Tielrooij, Paparo, Piatkowski, Bakker, and Bonn]{tielrooij2009dielectric}
Tielrooij,~K.; Paparo,~D.; Piatkowski,~L.; Bakker,~H.; Bonn,~M. Dielectric relaxation dynamics of water in model membranes probed by terahertz spectroscopy. \emph{Biophysical Journal} \textbf{2009}, \emph{97}, 2484--2492\relax
\mciteBstWouldAddEndPuncttrue
\mciteSetBstMidEndSepPunct{\mcitedefaultmidpunct}
{\mcitedefaultendpunct}{\mcitedefaultseppunct}\relax
\EndOfBibitem
\bibitem[Pegado \latin{et~al.}(2008)Pegado, J{\"o}nsson, and Wennerstr{\"o}m]{pegado2008like}
Pegado,~L.; J{\"o}nsson,~B.; Wennerstr{\"o}m,~H. Like-charge attraction in a slit system: pressure components for the primitive model and molecular solvent simulations. \emph{Journal of Physics: Condensed Matter} \textbf{2008}, \emph{20}, 494235\relax
\mciteBstWouldAddEndPuncttrue
\mciteSetBstMidEndSepPunct{\mcitedefaultmidpunct}
{\mcitedefaultendpunct}{\mcitedefaultseppunct}\relax
\EndOfBibitem
\bibitem[Conway \latin{et~al.}(1951)Conway, Bockris, and Ammar]{conway1951dielectric}
Conway,~B.; Bockris,~J.; Ammar,~I. The dielectric constant of the solution in the diffuse and Helmholtz double layers at a charged interface in aqueous solution. \emph{Transactions of the Faraday Society} \textbf{1951}, \emph{47}, 756--766\relax
\mciteBstWouldAddEndPuncttrue
\mciteSetBstMidEndSepPunct{\mcitedefaultmidpunct}
{\mcitedefaultendpunct}{\mcitedefaultseppunct}\relax
\EndOfBibitem
\end{mcitethebibliography}

\end{document}